\documentstyle[preprint,aps]{revtex}
  
\newcommand{\AmS}{{\protect\the\textfont2
  A\kern-.1667em\lower.5ex\hbox{M}\kern-.125emS}}
\newcommand{\be}{\begin{equation}}
\newcommand{\ee}{\end{equation}}
\newcommand{\ben}{\begin{eqnarray}}
\newcommand{\een}{\end{eqnarray}}
\newcommand{\nn}{\nonumber}
\newcommand{\op}{{\cal O}}
\newcommand{\logn}{\hspace{3pt}{\rm ln}}
\newcommand{\pole}[1]{{#1}/{\tilde \epsilon}}
\newcommand{\lnmup}{\hspace{3pt}{\rm ln}\left|{\mu^2}/{p^2}\right|}
\newcommand{\sla}[2]{{{#1}\hspace{-5pt}{/}}_{#2}}

\def\simgt{\rlap{\lower 3.5 pt\hbox{$\mathchar \sim$}}\raise 1pt \hbox {$>$}}
\def\simlt{\rlap{\lower 3.5 pt\hbox{$\mathchar \sim$}}\raise 1pt \hbox {$<$}}
\newcommand{\llop}{{\gamma_\mu^L\otimes\gamma_\mu^L}}
\newcommand{\llpp}{{\sla{p}{}^L\otimes\sla{p}{}^L/p^2}}

\begin{document}
\draft
\title{The Kaon $B$-parameter with the Wilson Quark 
       Action using Chiral Ward Identities}

\author{JLQCD Collaboration \\[1mm] 
        S.~Aoki$^{\rm a}$,
        M.~Fukugita$^{\rm b}$,
        S.~Hashimoto$^{\rm c}$,
        N.~Ishizuka$^{\rm a,d}$,
        Y.~Iwasaki$^{\rm a,d}$,
        K.~Kanaya$^{\rm a,d}$, 
        Y.~Kuramashi$^{\rm e}$
\thanks{On leave from Institute of Particle and Nuclear Studies,
High Energy Accelerator Research Organization(KEK),
Tsukuba, Ibaraki 305-0801, Japan},
        M.~Okawa$^{\rm f}$, 
        A.~Ukawa$^{\rm a,d}$,
        T.~Yoshi\'{e}$^{\rm a,d}$}

\address{$ ^a$Institute of Physics, 
         University of Tsukuba,\\ 
         Tsukuba, Ibaraki 305-8571, Japan\\
         $ ^b$Institute for Cosmic Ray Research, 
         University of Tokyo,\\ 
         Tanashi, Tokyo 188-8502, Japan\\
         $ ^c$Computing Research Center, 
         High Energy Accelerator Research Organization(KEK), \\
         Tsukuba, Ibaraki 305-0801, Japan\\
         $ ^d$Center for Computational Physics, 
         University of Tsukuba,\\ 
         Tsukuba, Ibaraki 305-8577, Japan\\
         $ ^e$Department of Physics, Washington University, 
         St. Louis, Missouri 63130, USA\\
         $ ^f$Institute of Particle and Nuclear Studies, 
         High Energy Accelerator Research Organization(KEK),\\ 
         Tsukuba, Ibaraki 305-0801, Japan}

\date{\today}

\maketitle

\newpage 

\begin{abstract}
A lattice QCD calculation of the kaon $B$ parameter $B_K$ is
carried out with the Wilson quark action 
in the quenched approximation at $\beta=6/g^2=5.9-6.5$.
The mixing problem
of the $\Delta s=2$ four-quark operators is solved non-perturbatively
with full use of chiral Ward identities
employing four external quarks with an equal off-shell
momentum in the Landau gauge. This method, without invoking
any effective theory, enables us to
construct the weak four-quark operators exhibiting good
chiral behavior.
Our results for $B_K$ with the non-perturbative mixing
coefficients show small scaling violation beyond the lattice
cut-off $a^{-1}\sim 2.5 $GeV.
Our estimate concludes $B_K({\rm NDR}, 2{\rm GeV})=0.69(7)$ at 
$a^{-1}=2.7-4.3$GeV,
which agrees with the value obtained with the Kogut-Susskind 
quark action.  
For comparison we also calculate $B_K$ with
one-loop perturbative mixing coefficients.  While this yields 
incorrect values at finite lattice spacing, 
a linear extrapolation to the continuum limit as a function
of $a$ leads to a result consistent with those obtained with
the Ward identity method.  

\end{abstract}


\section{Introduction}

The kaon $B$ parameter defined as a ratio
\be
B_K=\frac{\langle
{\bar K}^0 \vert \bar{s}\gamma_\mu(1-\gamma_5)d\cdot
\bar{s}\gamma_\mu(1-\gamma_5)d \vert K^0 \rangle}
{\frac{8}{3}\langle
{\bar K}^0 \vert \bar{s}\gamma_\mu\gamma_5 d\vert 0\rangle
\langle 0 \vert\bar{s}\gamma_\mu\gamma_5 d \vert K^0 \rangle}
\label{eq:defbk}
\ee
is one of the fundamental weak matrix elements
which have to be determined theoretically
for deducing the CP violation phase of the
Cabibbo-Kobayashi-Maskawa matrix from experiments.  
Lattice QCD calculation is expected to 
evaluate $B_K$ precisely incorporating the long-distance
effects of QCD. 
Much efforts have been devoted over the
years for this purpose, using both the Wilson and 
the Kogut-Susskind(KS) quark actions.
Successful calculations of $B_K$ have been
achieved so far with the KS
quark action, taking advantage of the correct chiral
behavior of the matrix element ensured 
by U(1) chiral symmetry\cite{bk_ks,saoki},
while studies with the Wilson quark action are rather stagnant.
There are two purposes for us to try to advance 
the calculations of $B_K$ with the Wilson action.
One of them is to verify the consistency between the Wilson and the
KS results, which would give a full credit to the lattice QCD
calculation. The other is an application 
to the heavy-light system for which   
the interpretation of flavor quantum numbers with the KS action is 
difficult.
   
An annoying defect of the Wilson quark action 
is explicit breaking of chiral symmetry 
at finite lattice spacing.
For the calculation of $B_K$ the problem 
appears as a non-trivial mixing of the weak $\Delta s=2$ four-quark operator 
of purely left handed chirality with those of mixed left-right 
chirality.  
Early studies showed that the mixing problem is not adequately
treated by perturbation theory, leading to an ``incorrect answer''
for the matrix element\cite{latt88}. Most calculations of $B_K$
have then tried to
solve the mixing problem non-perturbatively with the aid of 
chiral perturbation theory\cite{bk_w,z_ndr}, and have succeeded
in giving reasonable estimates for $B_K$.
This method,
however, is not promising from a point of view to control
systematic errors, since it contains large
uncertainties from higher order effects 
of chiral perturbation 
theory which survive even in the continuum limit.

An essential step toward a precise determination of $B_K$ 
is to control the operator mixing non-perturbatively
without resort to any effective theories.
The failure of the perturbative approach suggests that
higher order corrections in terms of the coupling
constant might be large in the mixing coefficients.  
Presence of large corrections in powers of the lattice spacing $a$
in the mixing coefficients is also a possibility.
In order to deal with this problem, the Rome group has proposed the
method of non-perturbative renormalization(NPR)\cite{npr}.  
Numerical results based on this approach show 
an improvement of the chiral behavior of the $\Delta s=2$
operator\cite{romeBK}.

In this paper we propose an alternative non-perturbative method 
to solve the operator mixing problem which is based on the use of 
chiral Ward identities\cite{wi}.
This method fully incorporates the chiral properties 
of the Wilson action explicitly.
We also reexamine the question
if perturbative mixing coefficients lead to erroneous
results for $B_K$ in the continuum limit.
Our simulations have been made within quenched QCD at
$\beta=5.9-6.5$ keeping the physical spatial size
approximately constant at $2.4$fm.
Chief findings of our calculation have already been presented 
in Ref.~\cite{bk_w_jlqcd} and we give in this article 
a detailed description of 
the implementation of our method and the results of our analyses.

This paper is organized as follows. 
In Sec.~\ref{sec:formulation} we describe
the formalism of our non-perturbative method to determine 
the mixing coefficients for four-quark operators 
based on the use of chiral Ward identities.
The perturbative expressions for the overall renormalization 
factors are also given.
In Sec.~\ref{sec:simulation} we present our data sets 
and give a description of the calculational 
procedure for $B_K$. Results for the mixing coefficients 
are given in Sec.~\ref{sec:result_zi}, 
where we compare our results
with those for the NPR method.
The overall renormalization
factors determined by the NPR method are compared
with those obtained by the perturbative one
in Sec.~\ref{sec:result_npr}.
In Sec.~\ref{sec:result_ch} we examine 
the chiral properties of the
four-quark operators constructed 
with the Ward identity method.
Final results for $B_K$ are presented in Sec.~\ref{sec:result_bk}.
Through Secs.~\ref{sec:result_zi}, \ref{sec:result_npr}, 
\ref{sec:result_ch}, and \ref{sec:result_bk} we also present results
with the perturbative method for comparative purposes.
Our conclusions are summarized in Sec.~\ref{sec:conclusion}.

\section{Formulation of the method}
\label{sec:formulation} 

\subsection{Determination of the mixing coefficients}
	
We first derive the generic form of the chiral Ward identities 
in a standard manner\cite{wi}.
The Wilson quark action is defined by
\ben
S_W&=&
-\frac{1}{2}\sum_{x,\mu}\left[\bar\psi(x)
(1-\gamma_\mu)U_\mu(x)\psi(x+\hat\mu)                
+\bar\psi(x+\hat\mu)(1+\gamma_\mu)U_\mu^\dagger(x)\psi(x)\right] \\ 
&&+\sum_x\bar\psi(x)(m_0+4)\psi(x)\nn,
\een
where $\psi=(u,d,s)$ represents the up, down and strange
quark fields.
The conventional hopping parameter is given by
\be
K=\frac{1}{2m_0+8}.
\ee
Under the flavor SU(3) chiral variation defined by
\ben
\delta^a \psi(x)&=&i\frac{\lambda^a}{2}\gamma_5\psi(x), \\
\delta^a {\bar \psi}(x)&=&{\bar \psi}(x)i\frac{\lambda^a}{2}\gamma_5, 
\een
with $\lambda^a$ $(a=1,\dots,8)$ the flavor matrices
normalized as Tr$(\lambda^a\lambda^b)=2\delta^{ab}$,
the naive Ward identity that follows from the Noether 
procedure takes the form:
\ben
&&\langle\nabla_\mu A_\mu^{{\rm ext},a}(x) 
\op(x_1,\cdots,x_n)\rangle\\ \nn
&=&2m_0\langle P^a(x) \op(x_1,\cdots,x_n)\rangle+
\langle X^a(x) \op(x_1,\cdots,x_n)\rangle\\ \nn
&&+i\langle\delta^a \op(x_1,\cdots,x_n)\rangle,
\een
where the pseudoscalar density $P^a$, the extended axial vector current 
$A_\mu^{{\rm ext},a}$ and its divergence are defined by
\ben
P^a(x)&=&{\bar \psi}(x)\frac{\lambda^a}{2}\gamma_5{\psi(x)},\\
A_\mu^{{\rm ext},a}(x)&=&\frac{1}{2}\left[\bar\psi(x)\frac{\lambda^a}{2}
\gamma_\mu\gamma_5 U_\mu(x)\psi(x+\hat\mu)
+\bar\psi(x+\hat\mu)\frac{\lambda^a}{2}
\gamma_\mu\gamma_5 U_\mu^\dagger(x)\psi(x)\right],\\
\nabla_\mu A_\mu^{{\rm ext},a}(x)&=&\sum_{\mu}\left[A_\mu^{{\rm ext},a}(x)
-A_\mu^{{\rm ext},a}(x-\hat\mu)\right],
\een
and the $X^a$ term is given by
\ben
X^a(x)&=&
-\frac{1}{2}\sum_\mu
\left[\bar\psi(x)\frac{\lambda^a}{2}\gamma_5 U_\mu(x)
\psi(x+\hat\mu)
+\bar\psi(x+\hat\mu)\frac{\lambda^a}{2}\gamma_5
U_\mu^\dagger (x)\psi(x)\right.\\ 
&&\left.+(x\rightarrow x-\hat\mu)\right]
+8\bar\psi(x)\frac{\lambda^a}{2}\gamma_5\psi(x).\nn
\een
The $X^a$ term mixes with $P^a$ and $\nabla_\mu A^a_\mu$ 
under renormalization. Thus we write
\be
X^a(x)=\bar X^a(x)-2\delta m_0 P^a(x)-(Z_A^{{\rm ext},a}-1)
\nabla_\mu A_\mu^{{\rm ext},a}(x),
\ee
where $\bar X^a(x)$ should satisfy:
\begin{itemize}
\item[(i)]On-shell matrix elements vanish in the continuum limit,
\be
\langle\alpha\vert\bar X^a(x)\vert\beta\rangle=O(a).
\ee
\item[(ii)]
Off-shell Green functions have only contact terms up to terms of $O(a)$,
\be
\langle \bar X^a(x) \op(x_1,\dots,x_n)\rangle=\sum_i\delta(x-x_i)
\langle{\op_i^{\prime\:a}}(x_1,\dots,x_n)\rangle+O(a).
\ee 
\end{itemize}
Defining the renormalized axial vector current by
\be
\hat A_\mu^{{\rm ext},a}(x)=Z_A^{\rm ext} A_\mu^{{\rm ext},a}(x),
\ee
and the renormalized quark mass by
\be
m=m_0-\delta m_0,
\ee
the Ward identity takes the following form:
\ben
&&\langle\nabla_\mu \hat A_\mu^{{\rm ext},a}(x) 
\op(x_1,\dots,x_n)\rangle\label{eq:wi}\\ 
&=&2m\langle P^a(x) \op(x_1,\dots,x_n)\rangle+
\langle \bar X^a(x) \op(x_1,\dots,x_n)\rangle\nn\\ 
&&+i\langle\delta^a \op(x_1,\dots,x_n)\rangle.\nn
\een
For finite quark masses it is also useful 
to take a four-dimensional sum over $x$, which gives
\ben
&&2m\sum_x \langle  P^a(x) \op(x_1,\dots,x_n)\rangle+
\sum_x\langle \bar X^a(x) \op(x_1,\dots,x_n)\rangle\label{eq:wisum}\\ 
&&+ \sum_x i\langle\delta^a \op(x_1,\dots,x_n)\rangle=0.\nn
\een
We note that $\bar X^a(x)$ in eqs.(\ref{eq:wi}) and (\ref{eq:wisum})
generates only contact terms up to terms of $O(a)$. 

Let us consider a set of weak operators in the continuum  
$\{\hat \op_i\}$ which closes under flavor 
chiral rotations 
$\delta^a \hat \op_i=ic^a_{ij}\hat \op_j$.  These operators are
given by linear combinations of a set of lattice local operators
$\{\op_\alpha\}$ as $\hat \op_i=\sum_\alpha Z_{i\alpha}\op_\alpha$.
We choose the mixing coefficients $Z_{i\alpha}$ such that
the Green functions of $\{\hat \op_i\}$ with quarks in
the external states satisfy the chiral
Ward identity to $O(a)$.  This identity is obtained from 
eq.(\ref{eq:wisum}):
\ben
&&-2\rho_m Z_A^{\rm ext}\sum_x \langle P^a(x)\hat \op_i(0)
\prod_k\tilde\psi(p_k)\rangle   
+c^a_{ij}\langle\hat \op_j(0)\prod_k\tilde\psi(p_k)\rangle 
\label{eq:wi_4qop}\\
&&-i\sum_l\langle\hat \op_i(0)\prod_{k\ne
l}\tilde\psi(p_k)\delta^a\tilde\psi(p_l)\rangle+O(a)=0,\nn 
\een
where $p_k$ is the momentum of the external quark.
We note that the first term in eq.(\ref{eq:wi_4qop}) 
comes from the chiral variation 
of the Wilson quark action and the third represents the chiral 
rotation of the external fields.
Since the identity (\ref{eq:wi_4qop}) is linear in the renormalization 
constants, the overall renormalization factor cannot be fixed.
Furthermore, with quarks in the external states, calculations have to 
be made in some fixed gauge, {\it e.g.,} the Landau gauge.

The $O(a)$ term is governed by the typical QCD scale
$\Lambda_{\rm QCD}$ at low external quark momenta,
while powers of $p_k a$ become the dominant 
source of cut-off effects as momenta increase.
To be able to impose the Ward identity to $O(a)$, 
we need to restrict the external momenta by 
the condition $p_k \ll 1/a$ for large momenta.  
On the other hand, no such bounds exist for small momenta
for the validity of the identity itself 
as long as $\Lambda_{\rm QCD}a\ll 1$.  

The parameter $\rho_m=(m_0-\delta m_0)/Z_A^{\rm ext}$ 
in eq.(\ref{eq:wi_4qop}) 
is determined from the PCAC(partial conservation of axial vector 
current) relation
obtained from an application of the Ward identity (\ref{eq:wi})
for $\op=P^b(y)$\cite{rhoza}:
\ben
\langle\nabla_\mu A_\mu^{{\rm ext},a}(x) P^b(y)\rangle
&=&2\rho_m {\langle P^a(x) P^b(y)\rangle}\label{eq:pcac}\\
&&-\delta (x-y)\frac{1}{Z_A^{\rm ext}}\langle{\bar\psi}(y) 
\left[\frac{1}{3}\delta^{ab}{\bf 1}
+d^{abc}\frac{\lambda^c}{2}\right]\psi(y)\rangle+O(a).\nn 
\een 
For the determination of $Z_A^{\rm ext}$
we employ another Ward identity:
\ben
-2\rho_m Z_A^{\rm ext}\sum_x \langle P^a(x) A_\mu^b(y) V_\nu^c(z)\rangle&=&
+if^{abd}\frac{Z_V}{Z_A}\langle V_\mu^d(y)V_\nu^c(z)\rangle \label{eq:wi_aav}\\ 
&&+if^{acd}\frac{Z_A}{Z_V}\langle A_\mu^b(y)A_\nu^d(z)\rangle+O(a),\nn
\een
where we take $\op=A_\mu^b(y) V_\nu^c(z)$ 
in eq.(\ref{eq:wisum})\cite{rhoza}.
The local axial vector current and the local vector current are defined by
\ben
A_\mu^b(y)&=&{\bar\psi(y)}\frac{\lambda^b}{2}
\gamma_\mu\gamma_5{\psi(y)},\\
V_\nu^c(z)&=&{\bar\psi(z)}\frac{\lambda^c}{2}
\gamma_\nu{\psi(z)},
\een
and $Z_A$ and $Z_V$ are the renormalization factors for
$A_\nu^b(y)$ and $V_\rho^c(z)$, respectively. 
The identity (\ref{eq:wi_aav}) can be regarded 
as a set of equations for $Z_A^{\rm ext}$ 
and $Z_A/Z_V$. Two independent equations are obtained from
$\mu=\nu=4$ and $\mu=\nu=i$ $(i=1,2,3)$.

The continuum four-quark operator relevant for $B_K$ is given by 
\be
\hat \op_{VV+AA}=(\bar s\gamma_\mu d)(\bar s\gamma_\mu d)
+(\bar s\gamma_\mu\gamma_5 d)(\bar s\gamma_\mu\gamma_5 d)
\label{eq:op_vv+aa}
\ee  
where $()$ means color trace, and the parity violating part of 
the operator which does not contribute to $B_K$ is dropped.
To fix the mixing coefficients for the lattice four-quark 
operators, we may choose a particular SU(3) flavor chiral
rotation to be applied for $\hat \op_{VV+AA}$. 
In order to avoid complexities in numerical simulations
it is essential to
avoid flavor rotations that yield operators
which have Penguin contractions and hence mix with lower dimension 
operators.
Assuming SU(2) symmetry $m_u=m_d$ 
we employ the $\lambda^3={\rm diag}(1,-1,0)$ chiral rotation,
under which
$\hat \op_{VV+AA}$ and  $\hat \op_{VA}=(\bar s\gamma_\mu d)
(\bar s\gamma_\mu\gamma_5 d)$ form a minimal closed set of 
the operators:
\ben
\delta^3\frac{1}{2}{\hat \op_{VV+AA}}=-i{\hat \op_{VA}}, \\
\delta^3{\hat \op_{VA}}=-i\frac{1}{2}{\hat \op_{VV+AA}}.
\een

Since $\hat \op_{VV+AA}$ and $\hat \op_{VA}$ are dimension six
operators with $\Delta s=2$, we can restrict ourselves to 
dimension six operators for the construction of the lattice
operators corresponding to them.
The set of lattice bare operators with even parity is given by  
\ben
VV&=&(\bar s\gamma_\mu d)(\bar s\gamma_\mu d), \\ 
AA&=&(\bar s\gamma_\mu\gamma_5 d)(\bar s\gamma_\mu\gamma_5 d), \\  
SS&=&(\bar s d)(\bar s d), \\ 
PP&=&(\bar s\gamma_5 d)(\bar s\gamma_5 d), \\ 
TT&=&\frac{1}{2}(\bar s\sigma_{\mu\nu} d)(\bar s\sigma_{\mu\nu} d), 
\een
and the set with odd parity is 
\ben
VA&=&(\bar s\gamma_\mu d)(\bar s\gamma_\mu\gamma_5 d), \\ 
SP&=&(\bar s d)(\bar s\gamma_5 d), \\  
T{\tilde T}&=&\frac{1}{2}
(\bar s\sigma_{\mu\nu} d)(\bar s\sigma_{\mu\nu}\gamma_5 d),  
\een
where $\sigma_{\mu\nu}=[\gamma_\mu, \gamma_\nu]/2$.  
We rearrange these operators into the Fierz eigenbasis,
which we find convenient 
when taking fermion contractions for evaluating the Green
functions in eq.(\ref{eq:wi_4qop}),
\ben
\op_0=& VV + AA &\;\;\;(+,+),
\label{eq:op_0} \\
\op_1=& SS+TT+PP &\;\;\;(+,+),
\label{eq:op_1} \\
\op_2=& SS-\frac{1}{3}TT+PP &\;\;\;(-,+), 
\label{eq:op_2}\\ 
\op_3=& \left(VV-AA\right) + 2\left(SS-PP\right) &\;\;\;(-,+), 
\label{eq:op_3}\\
\op_4=& \left(VV-AA\right) - 2\left(SS-PP\right) &\;\;\;(+,+), 
\label{eq:op_4}
\een
\ben
\op_5=& VA &\;\;\;(+,+), 
\label{eq:op_5}\\
\op_6=& SP+\frac{1}{2}T{\tilde T} &\;\;\;(+,-), 
\label{eq:op_6}\\
\op_7=& SP-\frac{1}{6}T{\tilde T} &\;\;\;(-,-). 
\label{eq:op_7}
\een
Here the first sign after each equation denotes the Fierz
eigenvalue and the second the $CPS$\cite{latt88} eigenvalue. 
The Fierz eigenbasis we employ is different from 
that chosen by the Rome group\cite{romeBK} based on one-loop
perturbation theory.

The parity odd operators $\op_{6,7}$ are $CPS$
odd while $\op_5$ is $CPS$ even,  and hence 
$\op_5$ does not mix
with $\op_{6,7}$ under renormalization, where we assume
$m_d=m_s$ in the quark action.
Therefore the mixing structure of these operators is given by
\ben
\frac{\hat \op_{VV+AA}}{Z_{VV+AA}}&=&\op_{VV+AA}
=z_0\op_0+z_1\op_1+ \cdots+z_4\op_4, 
\label{eq:mix_VV+AA} \\
\frac{\hat \op_{VA}}{Z_{VA}}&=&\op_{VA}=z_5\op_5,
\label{eq:mix_VA}
\een
where $Z_{VV+AA}$ and $Z_{VA}$ are overall renormalization factors, 
and we take $z_0=1$.

Let us consider an external state consisting of two $s$ quarks 
and two $d$ quarks, all having an equal momentum $p$.   
Under $\lambda^3$ chiral rotation the Ward identity (\ref{eq:wi_4qop}) 
for such an external state takes the following form:
\ben
F_{VV+AA}&\equiv& 
-2\rho_m Z_A^{\rm ext}\sum_x \langle 
P^a(x)\frac{1}{2}{\hat O}_{VV+AA}(0)
{\tilde s}(p){\tilde s}(p)
{\tilde {\bar d}}(p){\tilde {\bar d}}(p)\rangle 
\label{eq:f_vv+aa} \\ 
&&-\langle\hat \op_{VA}(0)
{\tilde s}(p){\tilde s}(p)
{\tilde {\bar d}}(p){\tilde {\bar d}}(p)\rangle \nn\\  
&&-\langle\frac{1}{2}{\hat O}_{VV+AA}(0)
{\tilde s}(p){\tilde s}(p)
\left[{\tilde {\bar d}}(p)\frac{\gamma_5}{2}\right]
{\tilde {\bar d}}(p)\rangle\nn \\  
&&-\langle\frac{1}{2}{\hat O}_{VV+AA}(0)
{\tilde s}(p){\tilde s}(p)
{\tilde {\bar d}}(p)
\left[{\tilde {\bar d}}(p)\frac{\gamma_5}{2}\right]\rangle+O(a)=0,\nn\\
F_{VA}&\equiv& 
-2\rho_m Z_A^{\rm ext}\sum_x \langle P^a(x){\hat O}_{VA}(0)
{\tilde s}(p){\tilde s}(p)
{\tilde {\bar d}}(p){\tilde {\bar d}}(p)\rangle 
\label{eq:f_va} \\ 
&&-\langle\frac{1}{2}{\hat O}_{VV+AA}(0)
{\tilde s}(p){\tilde s}(p)
{\tilde {\bar d}}(p){\tilde {\bar d}}(p)\rangle \nn \\  
&&-\langle{\hat O}_{VA}(0)
{\tilde s}(p){\tilde s}(p)
\left[{\tilde {\bar d}}(p)\frac{\gamma_5}{2}\right]
{\tilde {\bar d}}(p)\rangle\nn\\
&&-\langle{\hat O}_{VA}(0)
{\tilde s}(p){\tilde s}(p)
{\tilde {\bar d}}(p)
\left[{\tilde {\bar d}}(p)\frac{\gamma_5}{2}\right]\rangle
+O(a)=0.   \nn
\een
We obtain the amputated Green functions for 
$F_{VV+AA}$ and $F_{VA}$ by truncating the external quark
propagators according to 
\ben
\Gamma_{VV+AA}&\equiv& G_s^{-1}(p)G_s^{-1}(p)F_{VV+AA}
G_{\bar d}^{-1}(p)G_{\bar d}^{-1}(p),\\
\Gamma_{VA}&\equiv& G_s^{-1}(p)G_s^{-1}(p)F_{VA}
G_{\bar d}^{-1}(p)G_{\bar d}^{-1}(p),
\een
where $G_q^{-1}$ denotes the inverse quark propagator
with the flavor $q$. 

Let $P_i$ $(i=0,\dots,7)$ be the tree-level 
Dirac components  corresponding
to the four-quark operators in the Fierz eigenbasis 
$\op_i$ $(i=0,\dots,7)$,  {\it e.g.,}
\be
P_0^{\alpha\beta\delta\lambda}
=\gamma_\mu^{\alpha\beta}\gamma_\mu^{\delta\lambda}
+(\gamma_\mu\gamma_5)^{\alpha\beta}
(\gamma_\mu\gamma_5)^{\delta\lambda}.
\label{eq:dirac_0}
\ee
Since QCD conserves parity one can write
\ben 
\frac{\Gamma_{VV+AA}}{Z_{VV+AA}}&=&\Gamma_5P_5,\\
\frac{\Gamma_{VA}}{Z_{VA}}&=&\Gamma_0P_0
+\Gamma_1P_1+\cdots+\Gamma_4P_4,
\een
where the c-number coefficients 
$\Gamma_0,\dots,\Gamma_5$ are obtained by applying
the suitable projection operators to 
${\Gamma_{VV+AA}}/{Z_{VV+AA}}$ 
and ${\Gamma_{VA}}/{Z_{VA}}$, {\it e.g.,}
\be
{\overline P}_0^{\beta\alpha\lambda\delta}
=\frac{1}{128}
\left[\gamma_\nu^{\beta\alpha}\gamma_\nu^{\lambda\delta}
+(\gamma_5\gamma_\nu)^{\beta\alpha}
(\gamma_5\gamma_\nu)^{\lambda\delta}\right],
\label{eq:proj_0}
\ee
corresponding to $P_0^{\alpha\beta\delta\lambda}$.
Expressing ${\hat \op_{VV+AA,VA}}$ in eq.(\ref{eq:wi_4qop})
in terms of lattice operators,
we obtain six equations for the five coefficients $z_1,\dots,z_5$:
\be
\Gamma_i=c^i_0+c^i_1z_1+\cdots +c^i_5z_5=O(a),\, i=0,\dots, 5.
\ee 
This gives an overconstrained set of equations, and 
we may choose any five equations to exactly vanish to solve 
for $z_i$: the
remaining equation should automatically be satisfied to
$O(a)$. We choose four equations to be those for 
$i=1,\dots,4$, since $\op_1,\dots,\op_4$ do not
appear in the continuum.
The choice of the fifth equation, $i=0$ or 5,
is more arbitrary. We have checked that either $\Gamma_0=0$
or $\Gamma_5=0$ leads to a
consistent result to $O(a)$ for $z_1,\dots,z_4$ in the region
$pa\simlt 1$.
In the present analysis we choose $\Gamma_5=0$. 

Let us remark here that the equations obtained in the NPR
method \cite{romeBK} corresponds to $\Gamma_i=0$ for $i=1, \dots, 4$ 
in which the contributions of the first term due to the quark mass 
contributions and the third term representing chiral rotation of 
quark fields in the Ward 
identity (\ref{eq:wi_4qop}) are dropped.  
The authors of Refs.~\cite{npr,romeBK,marti_latt98} argued that
the NPR method is 
equivalent to the Ward identities  
in the limit of large external quark momentum $p$.  
The reasoning is that the first term is suppressed by one power of $p$
due to the explicit quark mass factor, and the third term
does not yield chiral-breaking components since 
the inverse quark propagator for large $p$ has the form 
$G^{-1}\propto i\sum_{\mu}\gamma_{\mu}p_{\mu}$ which
anticommutes with $\gamma_5$.
Under these circumstances 
the first, third and fourth terms of $F_{VA}$ 
in eq.(\ref{eq:f_va}) become
irrelevant for the determination of the
mixing coefficients $z_1,\dots,z_4$.
However, the latter point is not correct at finite
lattice spacing. 
The inverse quark propagator does not anticommute with
$\gamma_5$ in the large momentum region because
the contribution of the Wilson term in the quark propagator
becomes larger, and hence not negligible, as the momentum increases.
Therefore 
the third and fourth terms of $F_{VA}$ in eq.(\ref{eq:f_va})
yield components having Dirac structures other than $VV+AA$
after truncating the external quark propagators.
In conclusion the NPR method 
is a part of the Ward identities;
at the large external quark momentum $p$ 
the former becomes equivalent to the latter up to $O(pa)$.
In Sec.~\ref{sec:result_zi} we show the difference between
them numerically.

 
\subsection{Matching of lattice and continuum operators}

In our earlier report\cite{bk_w_jlqcd} we employed the NPR method 
of Ref.~\cite{npr} to evaluate the overall renormalization factor
$Z_{VV+AA}$ in eq.(\ref{eq:mix_VV+AA}).   For the reasons discussed 
in Sec.~\ref{sec:result_npr} we
use the perturbative estimate in the final analysis presented in 
this article. 

One-loop perturbative renormalization 
of the $\Delta_s=2$ operator is written 
in the following way\cite{pt_4}: 
\be
{\hat \op_{VV+AA}}
=Z_{VV+AA}\op_0
+\frac{\alpha_s}{4\pi}Z^*\left(\frac{1}{3}\op_1
-\frac{1}{2}\op_3-\frac{5}{12}\op_4\right),
\label{eq:z_mix_pt}
\ee
with $\op_i$ (i=0,\dots,4) being the Fierz eigen operators
defined in eqs.(\ref{eq:op_0})$-$(\ref{eq:op_4}).
Employing the $\overline{\rm MS}$ subtraction 
scheme with naive dimensional
regularization(NDR) for the continuum theory 
the renormalization factors are
given by\cite{pt_4,z_ndr}
\ben
Z_{VV+AA}&=&1+\frac{\alpha_s}{4\pi}
\left(-4\logn(\mu a)+\Delta_{VV+AA}\right),
\label{eq:z_vv+aa_pt}\\
\Delta_{VV+AA}&=&-50.841\hspace{24pt}{\rm for\hspace{12pt}NDR},  \\
Z^*&=&9.6431,
\een
where $\mu$ is the renormalization scale. 
The diagonal part $\Delta_{VV+AA}$ is affected by
the renormalization scheme in the continuum,
while the mixing part $Z^*$ is independent. 
We collect the value of $\Delta_{VV+AA}$ for the dimensional
reduction(DRED) scheme in the Appendix.

Including the normalization of quark fields 
$\sqrt{8K_c}\sqrt{1/2K-3/8K_c}$\cite{lm,kmac} 
tadpole-improved by the factor $u_0=(8K_c)^{-1}$ leads to
\ben
Z_{VV+AA}&=&\left(\frac{1}{2K}-\frac{3}{8K_c}\right)^2
\left[1+\frac{\alpha_s}{4\pi}
\left(-4\logn(\mu a)+\Delta_{VV+AA}
+2\pi\times 5.457 \right)\right],
\label{eq:z_vv+aa_pt_tad}
\een
Here $K_c$ is the critical hopping parameter 
where the pion mass vanishes.
We use 
\be
\frac{1}{8K_c}=1-5.457\alpha_s/4
\ee 
in Ref.\cite{pt_kc}
for the perturbative estimate of $K_c$.

With the use of $Z_{VV+AA}^{\rm NDR}$
we convert the matrix element on the lattice into 
that in the continuum NDR scheme renormalized at the scale
$\mu=1/a$GeV\cite{match,z_ndr}:
\be
B_K({\rm NDR},1/a)=\frac{Z_{VV+AA}^{\rm NDR}}{Z_A^{\rm NDR}}
\frac{\langle
{\bar K}^0 \vert \op_{VV+AA} \vert K^0 \rangle}
{\frac{8}{3}\vert \langle 0 \vert A_\mu \vert K^0 \rangle \vert^2}
\label{eq:Z-factor}
\ee
where $\op_{VV+AA}$ is defined in eq.(\ref{eq:mix_VV+AA}), and 
$Z_A$ is the renormalization factor for the axial vector
current $A_\mu={\bar s}\gamma_\mu\gamma_5 d$, which 
is expressed as\cite{pt_2,pt_kc}
\ben
Z_A&=&1+\frac{\alpha_s}{4\pi}\Delta_A,
\label{eq:z_a_pt}\\
\Delta_A&=&-21.061\hspace{24pt}{\rm for\hspace{12pt}NDR}. 
\een
The value of $\Delta_A$ for the 
DRED scheme is given in the Appendix.
With the tadpole-improvement the expression (\ref{eq:z_a_pt}) 
becomes
\ben
Z_A&=&\left(\frac{1}{2K}-\frac{3}{8K_c}\right)
\left[1+\frac{\alpha_s}{4\pi}\left(\Delta_A
+\pi\times 5.457\right)\right].
\label{eq:z_a_pt_tad}
\een

The continuum value at a physical scale $\mu=2$GeV is 
obtained via a two-loop renormalization group running
from $\mu=1/a$GeV:
\be
B_K({\rm NDR},\mu)=\left(\frac{\alpha_{\overline {\rm MS}}(\mu)}
{\alpha_{\overline {\rm MS}}(1/a)}\right)^{\frac{\gamma^{(0)}}{2\beta_0}}
\left[1+\frac{\alpha_{\overline {\rm MS}}(\mu)-
\alpha_{\overline {\rm MS}}(1/a)}{4\pi}
\left(\frac{\gamma^{(1)}\beta_0
-\gamma^{(0)}\beta_1}{2\beta_0^2}\right)\right]B_K({\rm NDR},1/a), 
\ee
where $\beta_{0,1}$ are the leading and next-to-leading
coefficients of the $\beta$ function and $\gamma^{(0,1)}$
are those of the anomalous dimension for ${\hat \op}_{VV+AA}$.
We take $\beta_0=11$, $\beta_1=102$, $\gamma^{(0)}=4$, 
and $\gamma^{(1)}=-7$\cite{bw} appropriate for the zero-flavor case
corresponding to our quenched calculation of $B_K$.
 
We define another $B$ parameter to investigate the chiral property
of the operator ${\hat{\op}}_{VV+AA}$:
\be
B_K^P({\rm NDR},1/a)=\frac{Z_{VV+AA}^{\rm NDR}}{Z_P^{\rm NDR}}
\frac{\langle
{\bar K}^0 \vert \op_{VV+AA} \vert K^0 \rangle}
{\frac{8}{3}\vert \langle 0 \vert P \vert K^0 \rangle \vert^2}
\label{eq:bk_p}
\ee
with $Z_P$ the renormalization factor 
for the pseudoscalar density $P={\bar s}\gamma_5 d$.
The continuum value of $B_K^P$ at $2$GeV is obtained by running 
from $\mu=1/a$GeV to $2$GeV according to
the two-loop renormalization group.  We use 
$\gamma_P^{(0)}=-8$ and $\gamma_P^{(1)}=-404/3$\cite{bw} 
for the leading and next-to-leading coefficients 
of the anomalous dimension of the pseudoscalar density 
in the zero-flavor case.
The one-loop perturbative expression  
for $Z_P$ with the tadpole improvement is given by\cite{pt_2,pt_kc}  
\ben
Z_P&=&\left(\frac{1}{2K}-\frac{3}{8K_c}\right)
\left[1+\frac{\alpha_s}{4\pi}\left(8\logn(\mu a)+\Delta_P
+\pi\times 5.457\right)\right],
\label{eq:z_p_pt}\\
\Delta_P&=&-30.128\hspace{24pt}{\rm for\hspace{12pt}NDR}. 
\een
The value of $\Delta_P$ for the 
DRED scheme is given in the Appendix.

The overall renormalization factors $Z_{VV+AA}$, 
$Z_A$, and $Z_P$ can be alternatively 
determined by the NPR method\cite{npr}.
The NPR method closely follows what is usually done in the
perturbative renormalization.
The vertex corrections are extracted from the amputated
Green functions for off-shell external quark states with momentum
$p$ in the Landau gauge according to 
\ben
&&G_s^{-1}(p)G_s^{-1}(p)
\langle\op_{VV+AA}(0) 
{\tilde s}(p){\tilde s}(p)
{\tilde {\bar d}}(p){\tilde {\bar d}}(p)\rangle   
G_{\bar d}^{-1}(p)G_{\bar d}^{-1}(p) 
=\Lambda_0(p)P_0+\cdots, \\
&&G_s^{-1}(p)
\langle A_\mu(0) 
{\tilde s}(p){\tilde {\bar d}}(p)\rangle G_{\bar d}^{-1}(p) 
=\Lambda_{\gamma_\mu \gamma_5}(p)P_{\gamma_\mu \gamma_5}+\cdots, \\
&&G_s^{-1}(p)
\langle P(0) 
{\tilde s}(p){\tilde {\bar d}}(p)\rangle G_{\bar d}^{-1}(p) 
=\Lambda_{\gamma_5}(p)P_{\gamma_5}+\cdots, 
\een
where $P$'s are the tree-level Dirac components with 
$P_0$ given in eq.(\ref{eq:dirac_0}) and 
$P_{{\gamma_\mu \gamma_5},{\gamma_5}}$ defined by 
\ben
P_{\gamma_\mu \gamma_5}^{\alpha\beta}
&=&(\gamma_\mu\gamma_5)^{\alpha\beta},\\
P_{\gamma_5}^{\alpha\beta}
&=&{\gamma_5}^{\alpha\beta}.
\een
We should note that the amputated Green functions for
the bilinear operators can have extra Dirac components
besides their tree-level ones, which originate from 
contribution of the higher dimensional operators.  
The quark wave-function renormalization factor $Z_q(p)$ is
extracted from the quark self energy:
\be
Z_q(p)=\frac{{\rm Tr}\left(-i\sum_\mu \gamma_\mu 
{\rm sin}(p_\mu)G_q^{-1}(p)\right)}
{12\sum_\mu {\rm sin}^2(p_\mu)},
\ee
where the trace is applied for the Dirac and color indices.
In terms of the vertex corrections and the wave-function
renormalization factor one calculates $Z_{VV+AA}$, $Z_A$, and $Z_P$
imposing the following conditions
\ben
&&Z_{VV+AA}(p)Z_q^{-2}(p)\Lambda_0(p)=1, \\
&&Z_A(p)Z_q^{-1}(p)\left[\frac{1}{4}\sum_\mu
\Lambda_{\gamma_\mu \gamma_5}(p)\right]=1, \\
&&Z_P(p)Z_q^{-1}(p)\Lambda_{\gamma_5}(p)=1. 
\een
This renormalization scheme is called 
the regularization independent(RI) scheme.
In this scheme the renormalization constants
depend on the external state and the gauge.
The perturbative values of the renormalization 
constants $\Delta_{VV+AA}$, $\Delta_A$, and $\Delta_P$
defined in 
eqs.(\ref{eq:z_vv+aa_pt}), (\ref{eq:z_a_pt}), 
and (\ref{eq:z_p_pt}) for the RI scheme are given in the Appendix.


\section{Details of numerical simulation}
\label{sec:simulation} 

\subsection{Data sets}
  
Our calculations are made with the Wilson quark action and the plaquette 
gauge action at $\beta=5.9-6.5$ in quenched QCD. 
Table~\ref{tab:runpara} summarizes our run parameters. 
Gauge configurations are generated with the 5-hit pseudo heat-bath
algorithm.  At each value of $\beta$ 
four values of the hopping parameter $K$ are adopted such that the physical
point for the $K$ meson can be interpolated.  
The critical hopping parameter $K_c$ is determined by
extrapolating results for $m_\pi^2$ at the four hopping parameters 
linearly in $1/2K$ to $m_\pi^2=0$.
We take the down and strange quarks to be degenerate.  
The value of half the strange quark mass $m_s a/2$ is then estimated 
from the experimental ratio $m_K/m_\rho=0.648$.

The inverse lattice spacing $a^{-1}$ is determined from the
$\rho$ meson mass $m_\rho=770$MeV. 
The physical size of lattice is chosen to be approximately 
constant at $La\approx 2.4$fm.
To calculate the perturbative renormalization factors, 
we employ the strong coupling constant at the scale $1/a$ 
in the $\overline{\rm MS}$ scheme, evaluated by a two-loop 
renormalization group running starting from   
$1/g^2_{\overline{\rm MS}}(\pi/a)=
\langle{\rm Tr}U_P\rangle/g^2_{latt}+0.0246$ with 
$\langle{\rm Tr}U_P\rangle$ the averaged value of the
plaquette.

In order to calculate the mixing
coefficients $z_i$ $(i=1,\dots,5)$ with the Ward identity
method and
the renormalization factors 
$Z_{VV+AA}$, $Z_A$, and $Z_P$ with the NPR method, 
the latter for purpose of comparison with the 
perturbative values,  
we prepare a set of 
external quark momenta $p^{(i)}=(p_x^{(i)},p_y^{(i)},
p_z^{(i)},p_t^{(i)})$ $(i=1,\dots,\sim 40)$. 
These momenta are 
chosen  recursively according to the condition
that the $(i+1)$-th momentum $p^{(i+1)}a$ is the minimum number
satisfying 
\ben
(p^{(i+1)}a)^2\ge \delta_{p^2}\cdot (p^{(i)}a)^2,\\
p_x^{(i+1)}\le p_y^{(i+1)}\le p_z^{(i+1)},
\een 
for a given value of the increment parameter $\delta_{p^2}$ 
starting with $p^{(1)}a=(0,0,0,2\pi/T)$ 
where $T$ denotes the temporal lattice size.
In the case of multiple choices for the $i+1$-st momentum 
we take the momentum that has the largest value of $p_t^{(i+1)}$.
The choice of the value of $\delta_{p^2}$ is listed in 
Table~\ref{tab:runpara}.
We employ the momentum having $p^{(*)}\approx 2$GeV among the $p^{(i)}$'s 
for the analysis of $B$ parameters.

We estimate errors by the single elimination 
jackknife procedure for all measured quantities
except for the extrapolation to the continuum limit as a
function of $a$. 

\subsection{Calculational procedure}

Our calculations are carried out in three steps. 
We first calculate  $m_\pi$, $m_\rho$, $\rho_m$ and 
$Z_A^{\rm ext}$ using the hadron Green functions.
For this purpose 
quark propagators are solved in the Landau gauge
for the point source located at the origin
with the periodic boundary condition imposed in all four directions.
Following eq.(\ref{eq:pcac}) we can extract 
the $\rho_m$ parameter from the ratio
\ben
\rho_m&=&\frac{1}{2}\left[\frac{\sum_{\vec x}\langle\nabla_4 
A_4^{{\rm ext},3}({\vec x},t)P^3({\vec 0},0)\rangle}
{2\sum_{\vec x}\langle P^3({\vec x},t)P^3({\vec 0},0)\rangle}
+(t\rightarrow T-t+1)\right]\\
&\longrightarrow& 
\frac{\langle0\vert\nabla_4 A_4^{{\rm ext},3}\vert \pi^3\rangle}
{2 \langle0\vert P^3\vert \pi^3\rangle}
\;\;\;\;\;\;\;\; 0\ll t \ll T-1,\nn
\een
by fitting a plateau as a function of $t$, 
where 
\ben
\nabla_4 A_4^{{\rm ext},3}({\vec x},t)&=&
A_4^{{\rm ext},3}({\vec x},t) - A_4^{{\rm ext},3}({\vec x},t-1), \\
A_4^{{\rm ext},3}({\vec x},t)
&=&\frac{1}{4}\left[{\bar u}(\vec{x},t)
\gamma_4\gamma_5 U_4(\vec{x},t)u(\vec{x},t+1)
+{\bar u}(\vec{x},t+1)\gamma_4\gamma_5
U_4^{\dagger}(\vec{x},t)u(\vec{x},t) \right.\\
&&\left.\;\;\;\;\;\;\;\;-(u\leftrightarrow d)\right], \nn \\
P^3({\vec x},t)&=&\frac{1}{2}\left[
{\bar u}(\vec{x},t)\gamma_5 u(\vec{x},t)-(u\leftrightarrow d)\right].
\een
In order to determine $Z_A^{\rm ext}$ we make a zero-momentum
projection in $y$ in eq.(\ref{eq:wi_aav}):
\ben
-2\rho_m Z_A^{{\rm ext},3}\sum_{x,\vec{y}}\langle
P^3(x)A_\mu^+(\vec{y},t)V_\nu^-({\vec 0},0)\rangle&=&   
\frac{Z_V}{Z_A}\sum_{\vec y}\langle V_\mu^+(\vec{y},t)
V_\nu^-({\vec 0},0)\rangle \label{eq:za_wi} \\
&&-\frac{Z_A}{Z_V}\sum_{\vec y}\langle A_\mu^+(\vec{y},t)
A_\nu^-({\vec 0},0)\rangle,\nn 
\een
where the flavor matrices $\lambda^+$ and $\lambda^-$ are
defined by
\be
\lambda^+= \left(\begin{array}{lll}
                 0,1,0 \\
                 0,0,0 \\
                 0,0,0 \\
                 \end{array} \right),\;\;\;
\lambda^-= \left(\begin{array}{lll}
                 0,0,0 \\
                 1,0,0 \\
                 0,0,0 \\
                 \end{array} \right).
\ee 
At each time slice $t$ we obtain $Z_A^{{\rm ext},3}$ and
$Z_A/Z_V$ from the two independent equations corresponding to 
the choices
$\mu=\nu=4$ and $\mu=\nu=i$ ($i=1,2,3$) in eq.(\ref{eq:za_wi}).  
In Table~\ref{tab:hmass} we summarize the values 
of $m_\pi$, $m_\rho$, $\rho_m$, and 
$Z_A^{\rm ext}$ for the four hopping parameters
at each $\beta$.  

In terms of $\rho_m$ and $Z_A^{\rm ext}$ 
we determine the mixing coefficients $z_i$ $(i=1,\dots,5)$
according to the Ward identity (\ref{eq:wi_4qop}).
The quark Green functions 
having finite space-time momenta are constructed with
the point source quark propagators in the Landau gauge.
For calculation of the first term 
in the Ward identity (\ref{eq:wi_4qop}), 
we employ the source method\cite{source} 
to insert the pseudoscalar density.

The $B_K$ parameter is extracted from the following ratio 
of the hadron-three point
function divided by the two-point functions: 
\ben
R_A(t)&=&\frac{\sum_{{\vec x},{\vec y},{\vec z}}\langle
\op_{{\bar K}^0}({\vec x},T-1)
{\hat{\op}}_{VV+AA}({\vec y},t)\op_{K^0}^\dagger({\vec z},0)\rangle }
{\frac{8}{3} 
\sum_{{\vec x},{\vec y}}\langle 
\op_{{\bar K}^0}({\vec x},T-1){\hat A}(\vec{y},t)\rangle
\sum_{{\vec y^\prime},{\vec z}}\langle {\hat A}({\vec y^\prime},t)
\op_{K^0}^\dagger({\vec z},0)\rangle} 
\label{eq:ratio} \\
&\longrightarrow& 
\frac{1}{L^3}B_K({\rm NDR},1/a)
\;\;\;\;\;\;\;\; 0\ll t \ll T-1,\nonumber
\een
where operators are defined by
\ben
\op_{K^0}({\vec x},t)&=&{\bar s}(\vec{x},t)\gamma_5 d(\vec{ x},t),\\
\op_{\bar K^0}({\vec x},t)&=&{\bar d}(\vec{x},t)\gamma_5 s(\vec{ x},t),\\
{\hat{\op}}_{VV+AA}({\vec x},t)&=&
\sum_{i=0}^{4}Z_{VV+AA} z_i \op_i(\vec{ x},t),\\
{\hat A}({\vec x},t)&=&Z_A {\bar s}(\vec{x},t)
\gamma_4 \gamma_5 d(\vec{ x},t),
\een
with $Z_{VV+AA}$ and $Z_A$ given in
eqs.(\ref{eq:z_vv+aa_pt_tad}) 
and (\ref{eq:z_a_pt_tad}).
The contribution of each operator $\op_i$
$(i=0,\cdots,4)$ to $B_K({\rm NDR},1/a)$ can be measured by the ratio
\ben
R^i_A(t)&=&\frac{\sum_{{\vec x},{\vec y},{\vec z}}
\langle \op_{{\bar K}^0}(\vec{x},T-1)
Z_{VV+AA}z_i{\op}_i(\vec{y},t)
\op_{K^0}^\dagger(\vec{z},0)\rangle }{\frac{8}{3} 
\sum_{{\vec x},{\vec y}}
\langle \op_{{\bar K}^0}(\vec{x},T-1){\hat A}(\vec{y},t)\rangle
\sum_{{\vec y^\prime},{\vec z}}
\langle {\hat A}(\vec{y^\prime},t)\op_{K^0}^\dagger(\vec{z},0)\rangle}  
\label{eq:ratio_i}\\
&\longrightarrow& \frac{1}{L^3}
\frac{\langle{\bar K}^0 \vert Z_{VV+AA}z_i{\op}_i \vert K^0
\rangle}{\frac{8}{3}\vert \langle 0 \vert {\hat A} 
\vert K^0 \rangle \vert^2}
\;\;\;\;\;\;\;\; 0\ll t \ll T-1.\nn
\een 
The sum of $R^i_A(t)$ ($i=0,\dots,4$) is equal to $R_A(t)$.
The parameter $B_K^P({\rm NDR},1/a)$ 
defined in eq.(\ref{eq:bk_p}) is obtained from the ratio
\ben
R_P(t)&=&\frac{\sum_{{\vec x},{\vec y},{\vec z}}\langle
\op_{{\bar K}^0}({\vec x},T-1)
{\hat{\op}}_{VV+AA}({\vec y},t)
\op_{K^0}^\dagger({\vec z},0)\rangle }{\frac{8}{3} 
\sum_{{\vec x},{\vec y}}
\langle \op_{{\bar K}^0}({\vec x},T-1) \hat P({\vec y},t)\rangle
\sum_{{\vec y^\prime},{\vec z}}
\langle \hat P({\vec y^\prime},t)\op_{K^0}^\dagger({\vec z},0)\rangle}  
\label{eq:ratio_p}\\
&\longrightarrow& \frac{1}{L^3}
B_K^P({\rm NDR},1/a)
\;\;\;\;\;\;\;\; 0\ll t \ll T-1,\nonumber
\een
where the renormalized pseudoscalar density is
\be
{\hat P}({\vec x},t)=Z_P{\bar s}(\vec{x},t)\gamma_5 d(\vec{x},t)
\ee
with $Z_P$ in eq.(\ref{eq:z_p_pt}).
For calculation of the ratios $R_A$, $R_A^i$, 
and $R_P$ we solve 
quark propagators without gauge fixing employing wall
sources placed at the edges of lattice where the Dirichlet boundary
condition is imposed in the time
direction. 

The value of $B_K$ obtained 
with the Ward identity method 
depends on the 
external quark momentum $p^{(i)}$ 
at which the mixing coefficients are evaluated. 
To investigate the
quark mass dependence and $a$ dependence of $B_K$
we employ the averaged value 
of $B_K$ over the five momenta from $p^{(*-2)}$ to $p^{(*+2)}$ 
where $p^{(*)}$ represents the momentum nearest to $2$GeV. 
We employ the same procedure for the analysis of $B_K^P$. 

\section{Results for mixing coefficients}
\label{sec:result_zi} 

In Fig.\ \ref{fig:zmix_63_ours} we plot 
a typical result for the mixing coefficients $z_i$ $(i=1,\dots,4)$ 
as a function of the external quark momenta for the case of 
$K=0.15034$ at $\beta=6.3$. 
In order to evaluate the mixing coefficients 
we need to choose a specific scale $p^{(*)}$ that satisfies 
the condition $p^{(*)}a\ll 1$ to avoid cut-off contaminations.
We observe that the mixing coefficients 
show only weak dependence over a wide momentum range  
$0.02 \simlt p^2a^2 \simlt 1.0$, albeit $z_1$ and
$z_2$ have large errors in the small momentum 
region $ p^2a^2 \simlt 0.1$. 
This enables us to evaluate the mixing coefficients with small 
uncertainties from the choice of the momentum $p^{(*)}$. 
We adopt the value $p^{(*)}\approx 2$GeV, which we find to
always fall within the range of a plateau 
for our runs at $\beta=5.9-6.5$.

Let us compare the mixing coefficients obtained by the
Ward identity (WI) method with those by the NPR.
Since the NPR method does not employ the full Ward identity
of eq.(\ref{eq:wi_4qop}),
it is important to investigate differences in the
mixing coefficients between the NPR
and the Ward identity methods.
In Fig.\ \ref{fig:zmix_63_rome} we present
the result for the mixing coefficients $z_i$ obtained with
the NPR method.
The NPR result shows a strong scale dependence 
in the region $p^2 a^2\simlt
0.3$, which contrasts to the Ward identity result 
in Fig.\ \ref{fig:zmix_63_ours}. 
We suspect that this behavior of the NPR result originates
from physical non-perturbative contributions, 
which survive even 
in the continuum limit(see also Sec.~\ref{sec:result_npr}). 
Although we observe a similar scale dependence for the two results 
beyond the scale $p^2 a^2\sim 0.3$, 
the numerical values for $z_2$ and $z_3$
show clear deviations beyond the error bars 
between the WI result and that from the NPR
for momenta as large as $p^2 a^2\sim 2$. 
This is contrary to the expectation 
that the NPR would be equivalent to the Ward identities 
in the limit of large external 
quark momenta\cite{npr,romeBK,marti_latt98}. 

The difference between the NPR and the WI methods 
comes from the first and third terms in eq.(\ref{eq:wi_4qop}).
To investigate the contribution of each term to the mixing
coefficients
we reevaluate the mixing coefficients using the Ward
identities without the first term or the third term.
The former result is plotted in 
Fig.\ \ref{fig:zmix_63_1234_m50}(a) and the latter one in
Fig.\ \ref{fig:zmix_63_1234_m50}(b).
Comparison between Fig.\ \ref{fig:zmix_63_1234_m50}(a) and
Fig.\ \ref{fig:zmix_63_ours} demonstrates
that the contribution of the first term to the
mixing coefficients
is remarkable in the lower momentum region $p^2 a^2\simlt 0.3$.
Above this scale, the first term seems to play 
a minor role on the determination of the mixing coefficients.
On the other hand, 
comparing Fig.\ \ref{fig:zmix_63_1234_m50}(b) with
Fig.\ \ref{fig:zmix_63_ours} 
the essential contribution of the
third term to the mixing coefficients is observed 
over a wide range of external momentum, even up to
$p^2 a^2\sim 2$.   
In the Ward identity method
we can neglect
neither the first term nor the third one.

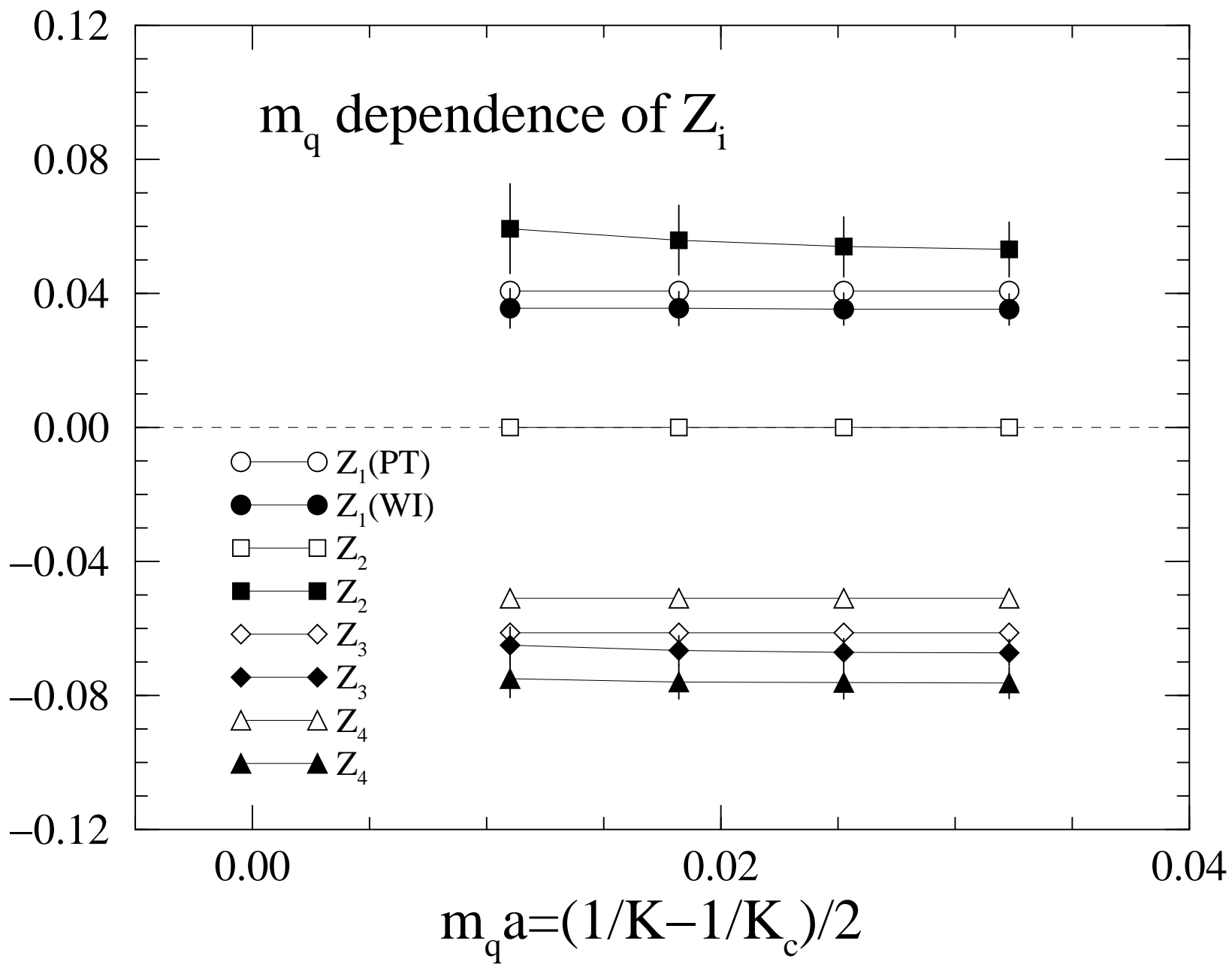
\begin{figure}
\caption{Quark mass dependence of mixing coefficients
$z_1,\dots,z_4$ evaluated at ${p^{(*)}}\approx 2$ GeV 
using the  Ward identity (WI; filled symbols) method 
at $\beta=6.3$.
Perturbative (PT; open symbols) results are also plotted for 
comparison.}  
\label{fig:zmix_63_1234_chl}
\end{figure}

Figure\ \ref{fig:zmix_63_1234_chl}
shows the quark mass dependence of the mixing
coefficients $z_i$ $(i=1,\dots,4)$ evaluated 
at the scale $p^{(*)}$ (filled symbols) for the case of 
$\beta=6.3$. For comparison we also plot
the perturbative (PT) estimate for $z_i$ (open symbols),
which are given in eq.(\ref{eq:z_mix_pt}), {\it i.e.,}
\ben
z_1&=&\frac{\alpha_s}{4\pi}Z^*\left(+\frac{1}{3}\right),\\
z_2&=&0, \\
z_3&=&\frac{\alpha_s}{4\pi}Z^*\left(-\frac{1}{2}\right),\\
z_4&=&\frac{\alpha_s}{4\pi}Z^*\left(-\frac{5}{12}\right),
\een
where $\alpha_{\overline{\rm MS}}(1/a)$ is used for the 
strong coupling constant. 
We observe little quark mass dependence for
the mixing coefficients. 

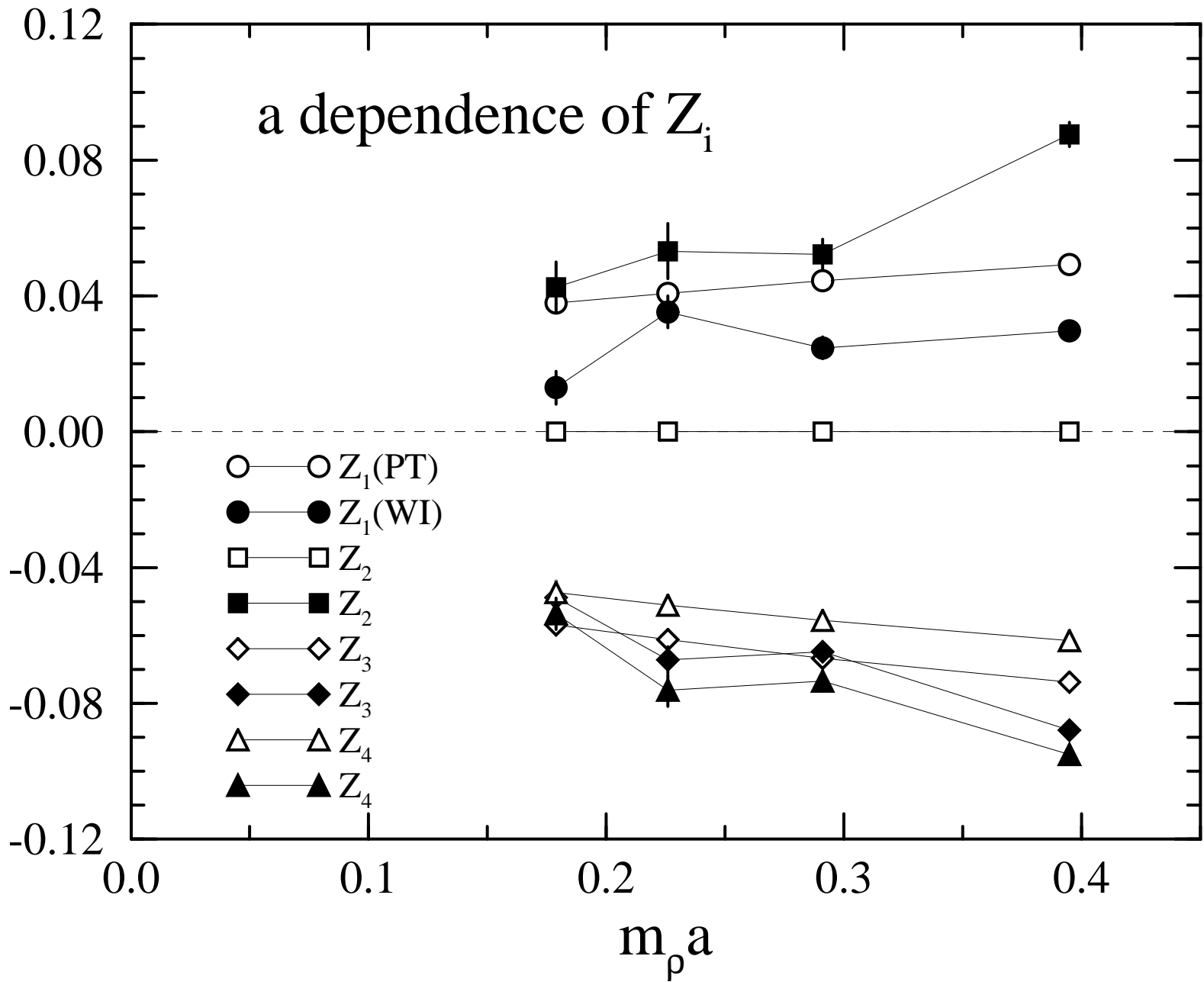
\begin{figure}
\caption{Comparison of mixing coefficients
$z_1,\dots,z_4$ evaluated at ${p^{(*)}}\approx 2$ GeV 
using the Ward identity (WI; filled symbols) method 
and the perturbative (PT; open symbols) one
as a function of $m_\rho a$.}  
\label{fig:zmix_1234_beta}
\end{figure}

In Fig.\ \ref{fig:zmix_1234_beta} we present the $a$ dependence
of the mixing
coefficients $z_i$ $(i=1,\dots,4)$ evaluated 
at the scale $p^{(*)}$ employing the heaviest quark mass
at each $\beta$.
We observe that the $a$ dependence of the mixing
coefficients determined by the Ward identities is steeper compared
to that of the PT estimates to one-loop order.
The magnitude of each mixing coefficient 
for the WI method varies nearly in
proportion to $a$, which reduces by $50\%$ between
$m_\rho a\approx 0.4$ and $m_\rho a\approx 0.2$.
A possible source of this $a$ dependence of the mixing
coefficients is the $O(a)$ term in
eq.(\ref{eq:wi_4qop}): contributions of the $O(a)$ term are absorbed
in the mixing coefficients to satisfy the continuum Ward
identities at finite lattice spacing.    

Comparing the mixing coefficients for the Ward identity method
and those of perturbation theory in 
Figs.\ \ref{fig:zmix_63_1234_chl} and \ \ref{fig:zmix_1234_beta},
we note that a large value of $z_2$ determined by the Ward identities 
sharply contrasts with the one-loop perturbative result $z_2=0$.  
The magnitude of this discrepancy appears larger than that possibly 
explained by two-loop contributions; 
squaring a typical magnitude of one-loop terms 
in Fig.~\ref{fig:zmix_1234_beta} only yields 
a value of order $\sim 0.001$. 
Discrepancies also exist for the other coefficients, albeit less 
conspicuous in that the PT results agree with those of 
WI in sign and
rough orders of magnitude.  In particular
the magnitude of $z_4$ is larger than that for 
$z_3$ for all values of $\beta$, which is contrary to the
perturbative result.

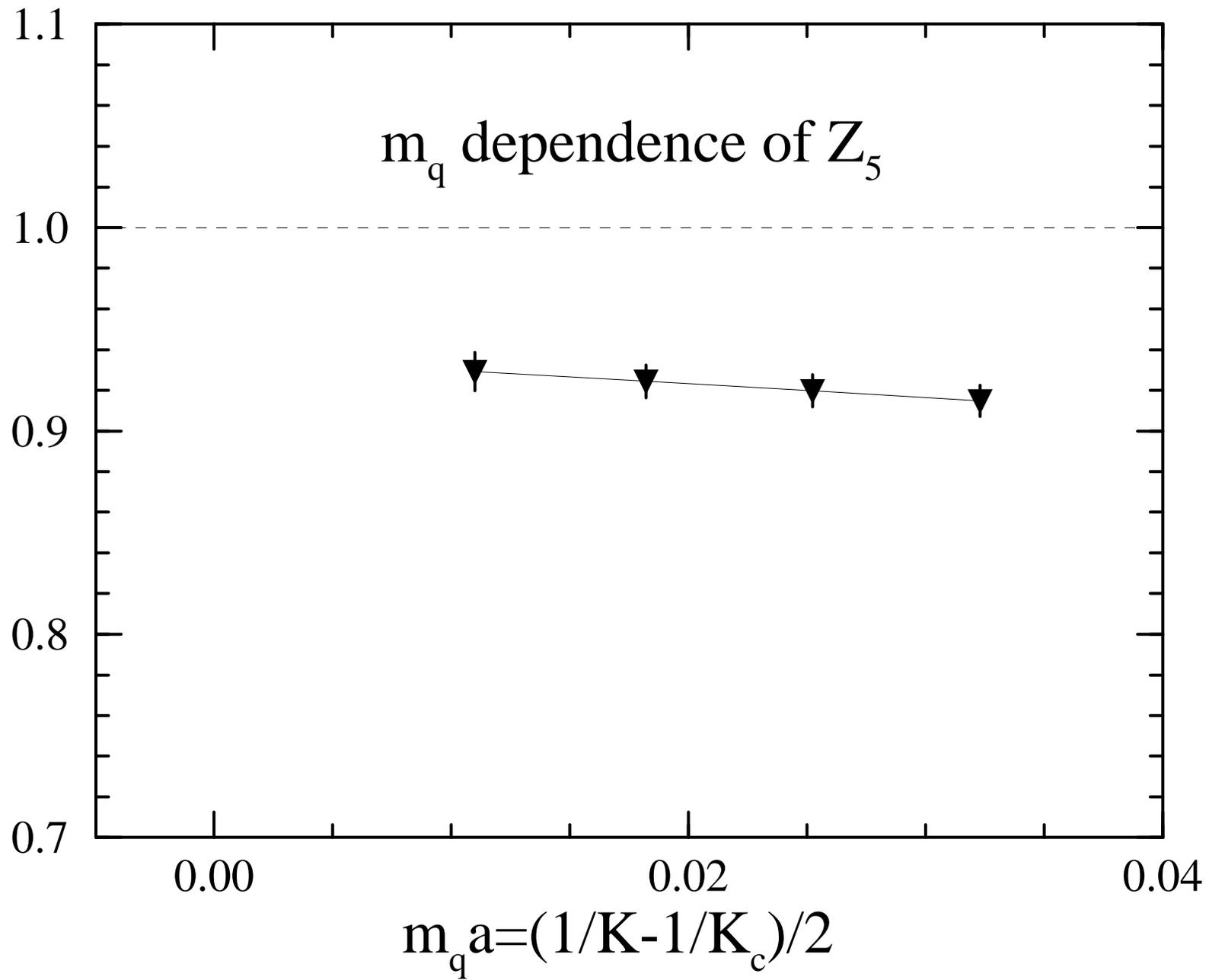
\begin{figure}
\caption{Quark mass dependence of mixing coefficient
$z_5$ evaluated at ${p^{(*)}}\approx 2$ GeV 
using the Ward identity method 
at $\beta=6.3$.} 
\label{fig:zmix_63_5_chl}
\end{figure}

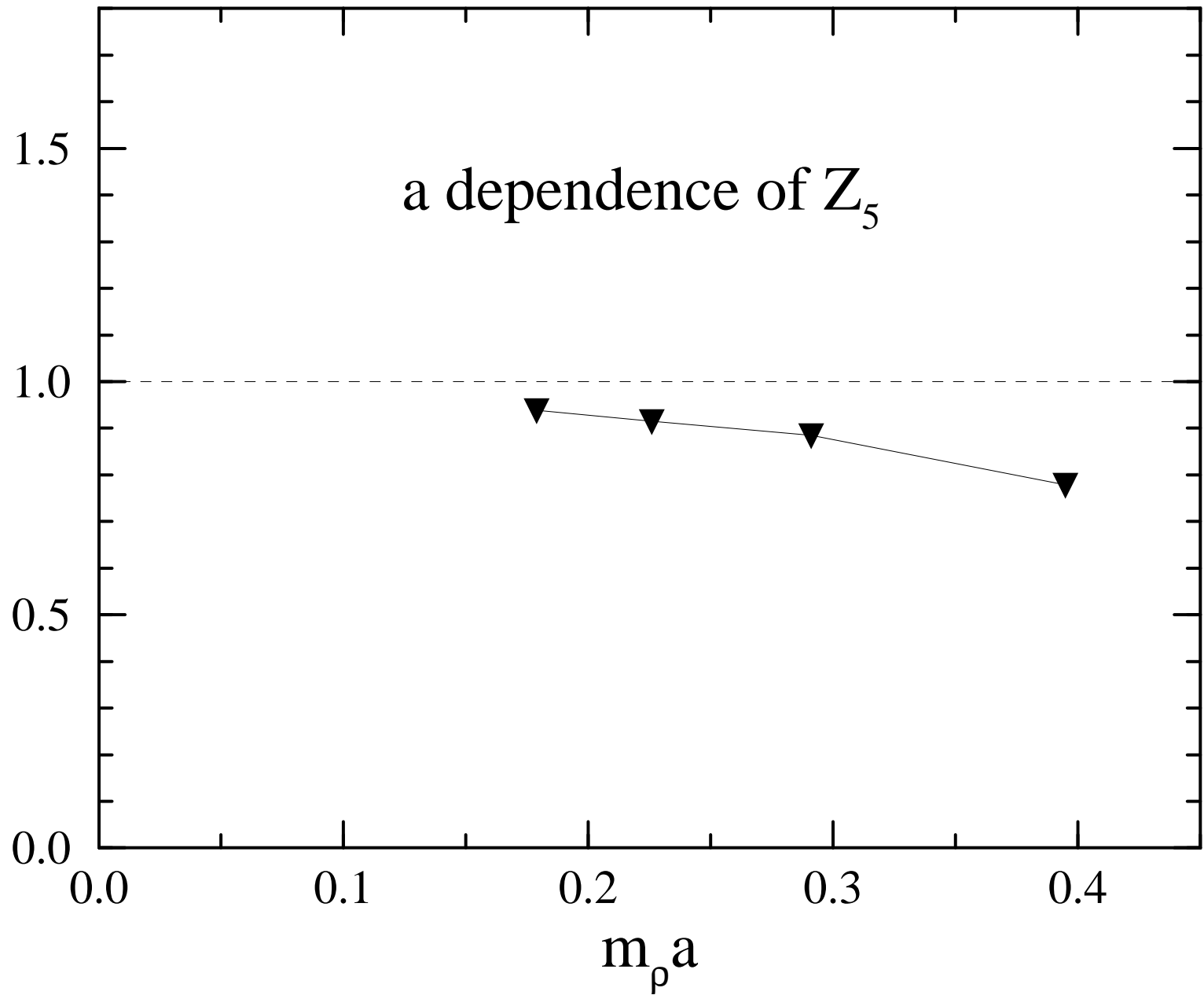
\begin{figure}
\caption{Lattice spacing dependence of 
mixing coefficient $z_5$ evaluated 
at ${p^{(*)}}\approx 2$ GeV using the Ward identity method.}  
\label{fig:zmix_5_beta}
\end{figure}

For our study of the $B$ parameter the mixing coefficient $z_5$
for the parity-odd operator ${\hat \op_{VA}}$ is not directly relevant. 
However, it is instructive to examine the scale
dependence of $z_5$, because it would take the value $z_5=1$ in 
the absence of cut-off dependent chiral symmetry breaking effects. 
In Fig.\ \ref{fig:zmix_63_5}
we plot a typical result for $z_5$. 
We find a scale dependence stronger than those of 
$z_i$ $(i=1,\dots,4)$ for parity-even operators toward large
momenta; the value of $z_5$ significantly deviates from unity as the
momentum increases, which measures the magnitude of cut-off effects.
The quark mass dependence of $z_5$ evaluated at $p^{(*)}$ is 
shown in Fig.\ \ref{fig:zmix_63_5_chl}. 
The value of $z_5$ slightly increases as the quark mass decreases.
We do not consider the strong scale dependence of $z_5$ to be particularly 
alarming since $z_5$ evaluated at a fixed 
physical scale $p^{(*)}$ approaches unity toward the continuum limit 
as shown in Fig.\ \ref{fig:zmix_5_beta}.

\section{Results for overall renormalization factors with
the NPR method}
\label{sec:result_npr} 
 
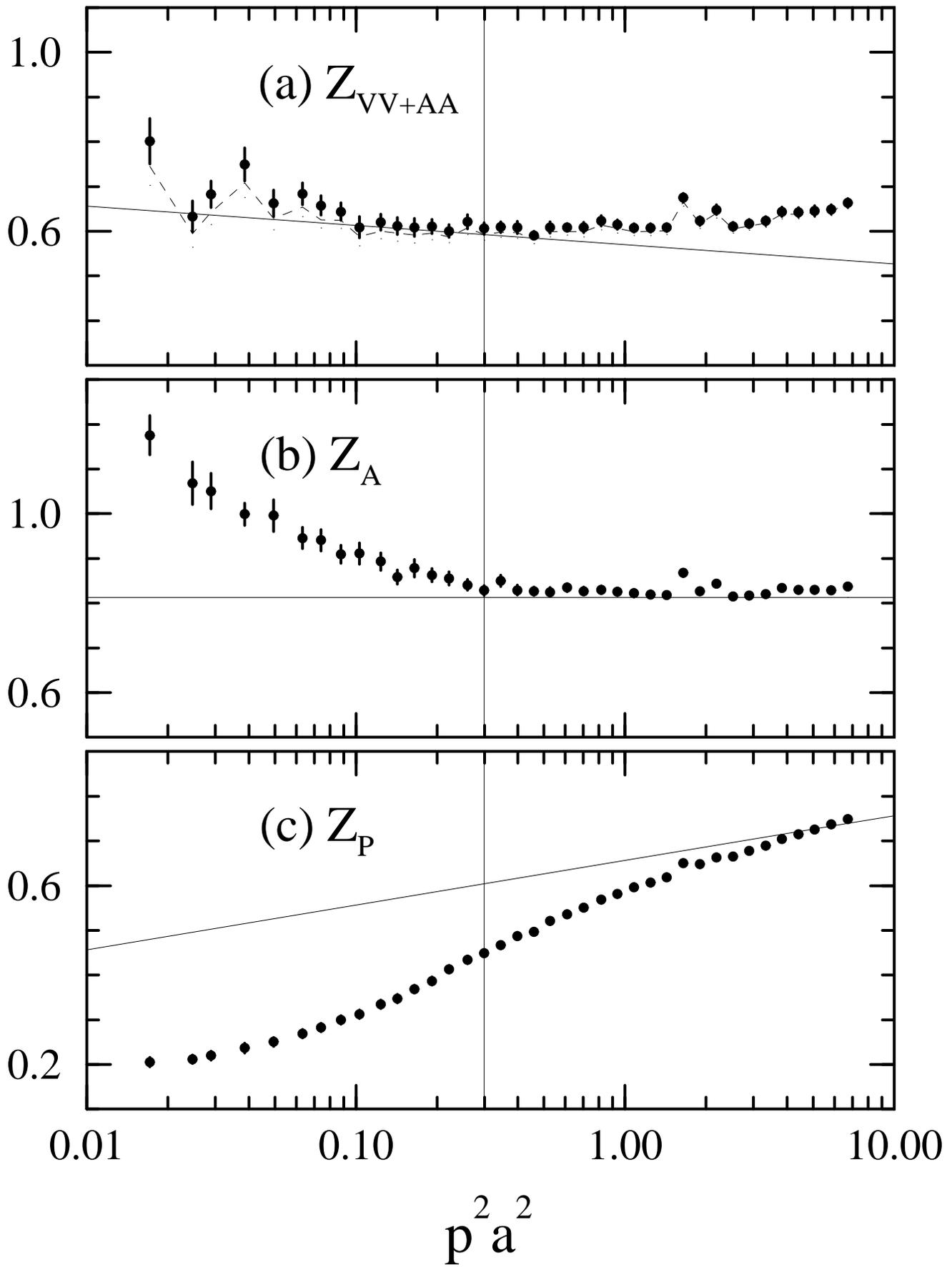
\begin{figure}
\caption{Renormalization factors (a)$Z_{VV+AA}$,
(b)$Z_A$, and (c)$Z_P$ in the RI scheme 
obtained with the NPR method
as a function of external momentum squared $p^2a^2$ 
for $K=0.15034$ at $\beta=6.3$.
Transverse lines denote tadpole-improved 
perturbative estimates. 
Dotted curve in (a) represents the $\op_0$ contribution
to $Z_{VV+AA}$.
Vertical lines correspond to 
$p^{(*)}\approx 2$ GeV.}
\label{fig:znpr_63}
\end{figure}

The NPR method is a possible way to estimate the overall
renormalization factors $Z_{VV+AA}$, $Z_A$, and $Z_P$.
In Fig.\ \ref{fig:znpr_63} we plot 
$Z_{VV+AA}$, $Z_A$, and $Z_P$ in the RI scheme  
as a function of $p^2 a^2$ for the
case of $K=0.15034$ at $\beta=6.3$.
For comparison we also draw 
the tadpole-improved one-loop perturbative estimates (solid lines) 
in the RI scheme.
The NPR result for $Z_{VV+AA}$ in Fig.\ \ref{fig:znpr_63}(a)
shows an agreement with the perturbative estimate
in the region $0.1\simlt p^2 a^2 \simlt 0.5$. 
The dotted curve in Fig.\ \ref{fig:znpr_63}(a) represents
the $\op_0$ contribution to $Z_{VV+AA}$, which is 
obtained by neglecting contributions of the mixed operators
$\op_i$ ($i=1,\dots,4$).
We observe that the contributions of the mixed operators,
leading of which is the two-loop radiative corrections, 
are quite small.

Figure\ \ref{fig:znpr_63}(b) shows that 
$Z_A$ has a strong 
$p^2 a^2$ dependence below $p^2 a^2 \sim 0.3$.
This behavior is contrary to the expectation that 
$Z_A$ should be independent of $p^2a^2$ as 
the axial vector current has no anomalous dimension. 
The unexpected behavior may be ascribed to  
non-perturbative contaminations due to 
the pion pole which could give an important
contribution at low external quark momentum\cite{npr}. 
In Fig.\ \ref{fig:znpr_63}(c) we observe
a large deviation between the NPR result for $Z_P$ 
and that from perturbation theory below $p^2 a^2 \sim 1$.
We suspect that this discrepancy is also due to 
the non-perturbative effects from the pion pole.
It should be remarked that the contamination due to the
pion pole is not a lattice artifact,  and hence 
survives even after taking the continuum limit.  

We note that our NPR results for $Z_A$ and $Z_P$ are 
consistent with those of recent studies\cite{gimenez,pittori,ishizuka}.
In particular evidence for the existence of pion pole contribution
in $Z_P$ has been reported in Ref.~\cite{ishizuka} for the Kogut-Susskind
action and in Ref.~\cite{pittori} for the Wilson case. 
 
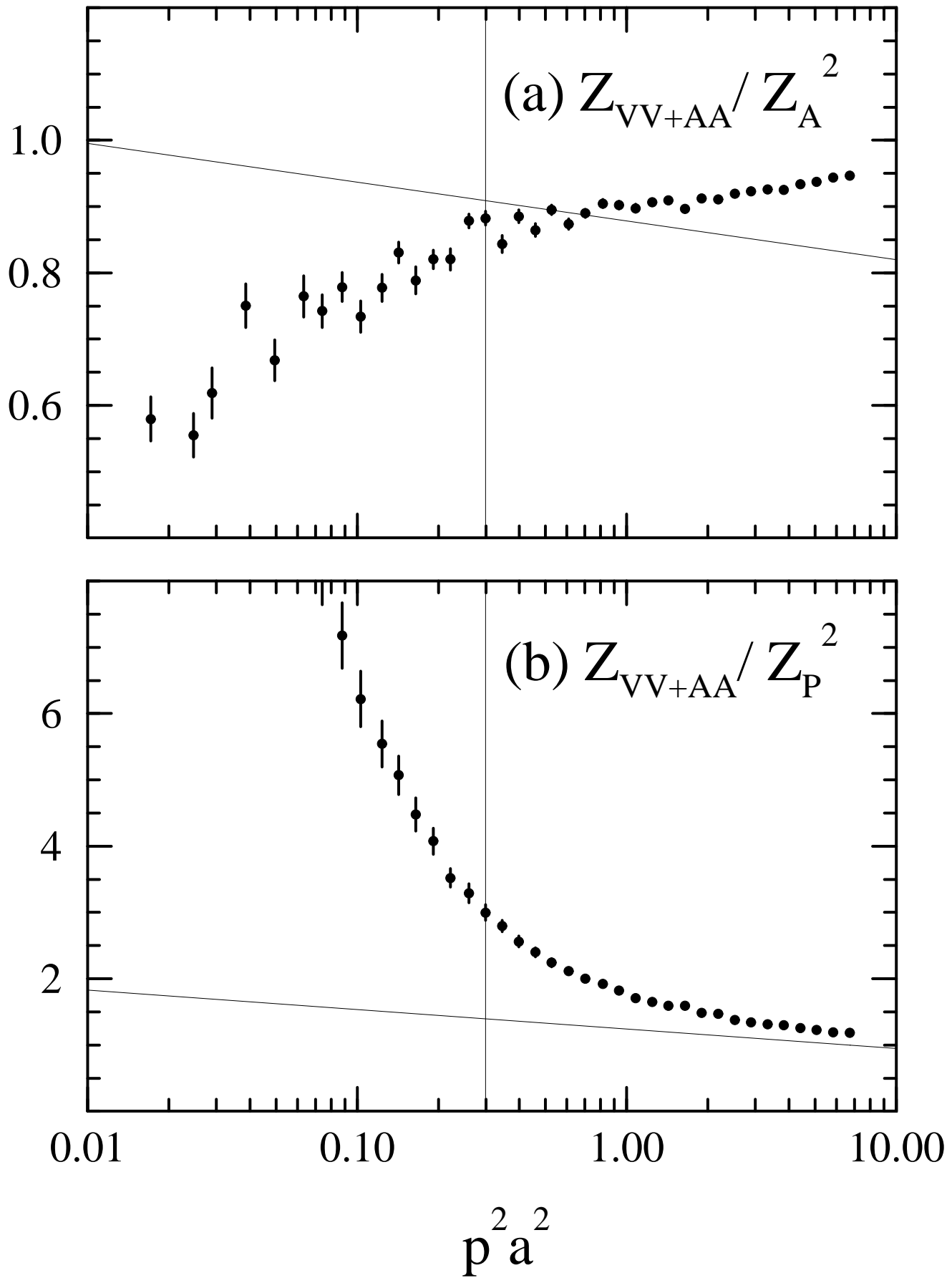
\begin{figure}
\caption{Ratios of renormalization factors 
(a)$Z_{VV+AA}/Z_A^2$ and (b)$Z_{VV+AA}/Z_P^2$
obtained with the NPR method
as a function of external momentum squared $p^2a^2$ 
for $K=0.15034$ at $\beta=6.3$.
Transverse lines denote tadpole-improved 
perturbative estimates. 
Vertical lines correspond to 
$p^{(*)}\approx 2$ GeV.}
\label{fig:znpr_63_r}
\end{figure}

For the calculation of $B$ parameters the ratios
$Z_{VV+AA}/Z_A^2$ and $Z_{VV+AA}/Z_P^2$ are more 
relevant. We show the scale dependence of 
$Z_{VV+AA}/Z_A^2$ in Fig.\ \ref{fig:znpr_63_r}(a)
and that of $Z_{VV+AA}/Z_P^2$ in Fig.\ \ref{fig:znpr_63_r}(b).
Solid lines are tadpole-improved one-loop results in the RI scheme. 
We observe that the NPR result for $Z_{VV+AA}/Z_A^2$
has a $p^2a^2$ dependence opposite to that expected
from the perturbative estimate. For the ratio 
$Z_{VV+AA}/Z_P^2$ the NPR result 
diverges toward lower momentum.

Application of the NPR method requires the existence of a
region $\Lambda_{\rm QCD}\ll p \ll 1/a$ where we can 
keep under control both non-perturbative
contaminations and cut-off effects, 
which is called a ``window'' in Ref.\cite{npr}. 
We find for $Z_{VV+AA}$, $Z_A$, $Z_P$, and their ratios 
that the lower bound of the window depends strongly on each
operator; we observe $p^2a^2\sim 0.1$ for $Z_{VV+AA}$ and 
$p^2a^2\sim 0.3$ for $Z_A$, while it is difficult to find  
the lower bound 
for $Z_P$, $Z_{VV+AA}/Z_A^2$, and $Z_{VV+AA}/Z_P^2$.
It is not clear to what extent 
non-perturbative contaminations can be separated out quantitatively. 
For these reasons we employ the perturbative estimates 
for $Z_{VV+AA}$, $Z_A$, and 
$Z_P$, rather than those of the NPR method, 
to obtain the $B$ parameter in our final analysis presented 
in this article. 
Numerical values of the $B_K$ parameter are little affected by this
change since the the ratio $Z_{VV+AA}/Z_A^2$ has a similar value 
at $p^{(*)}$ among the two methods. 
 
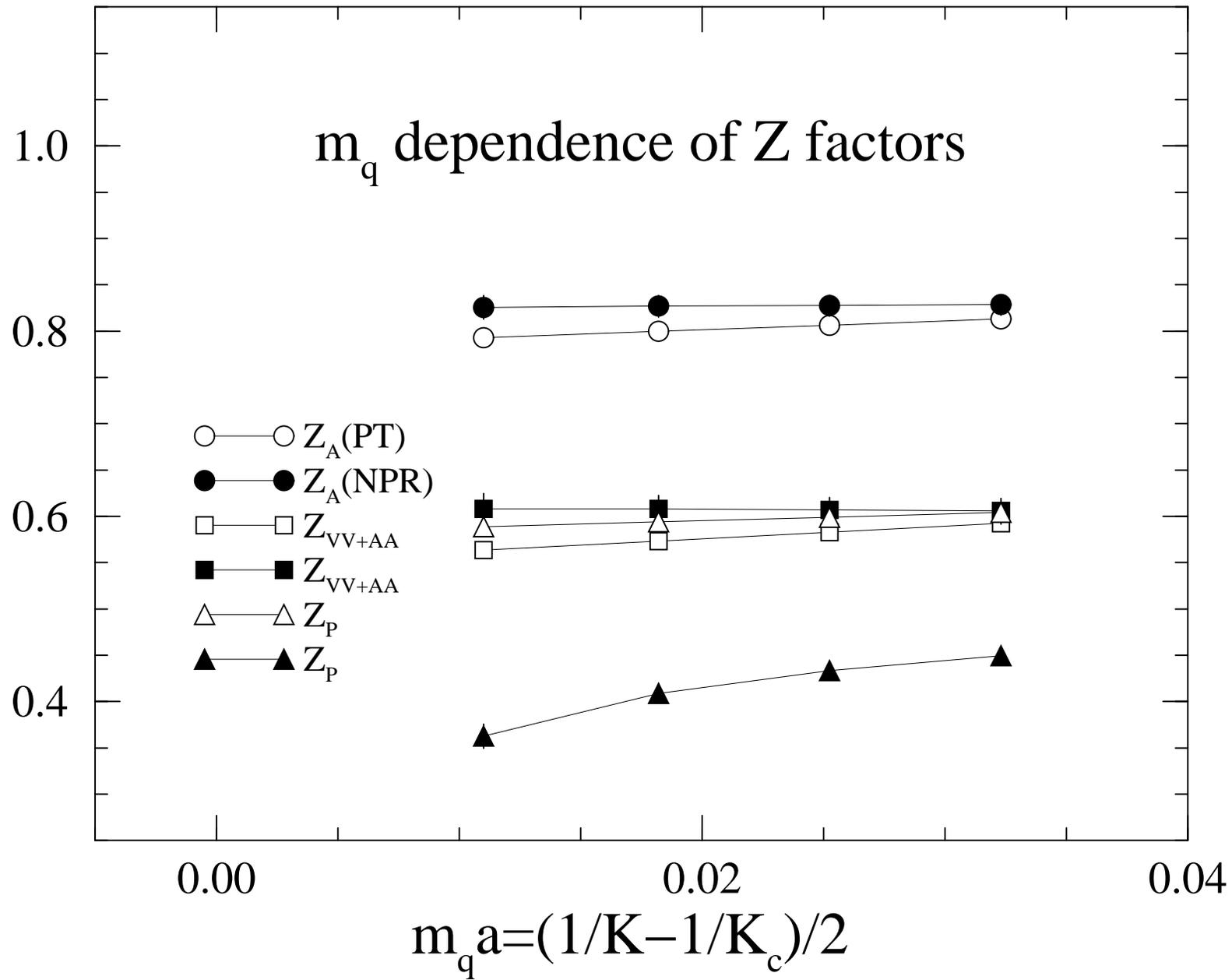
\begin{figure}
\caption{Quark mass dependence of renormalization factors
at ${p^{(*)}}\approx 2$ GeV 
using the NPR (solid symbols) method 
at $\beta=6.3$.
Open symbols represent tadpole-improved 
perturbative estimates.}  
\label{fig:znpr_63_chl}
\end{figure}

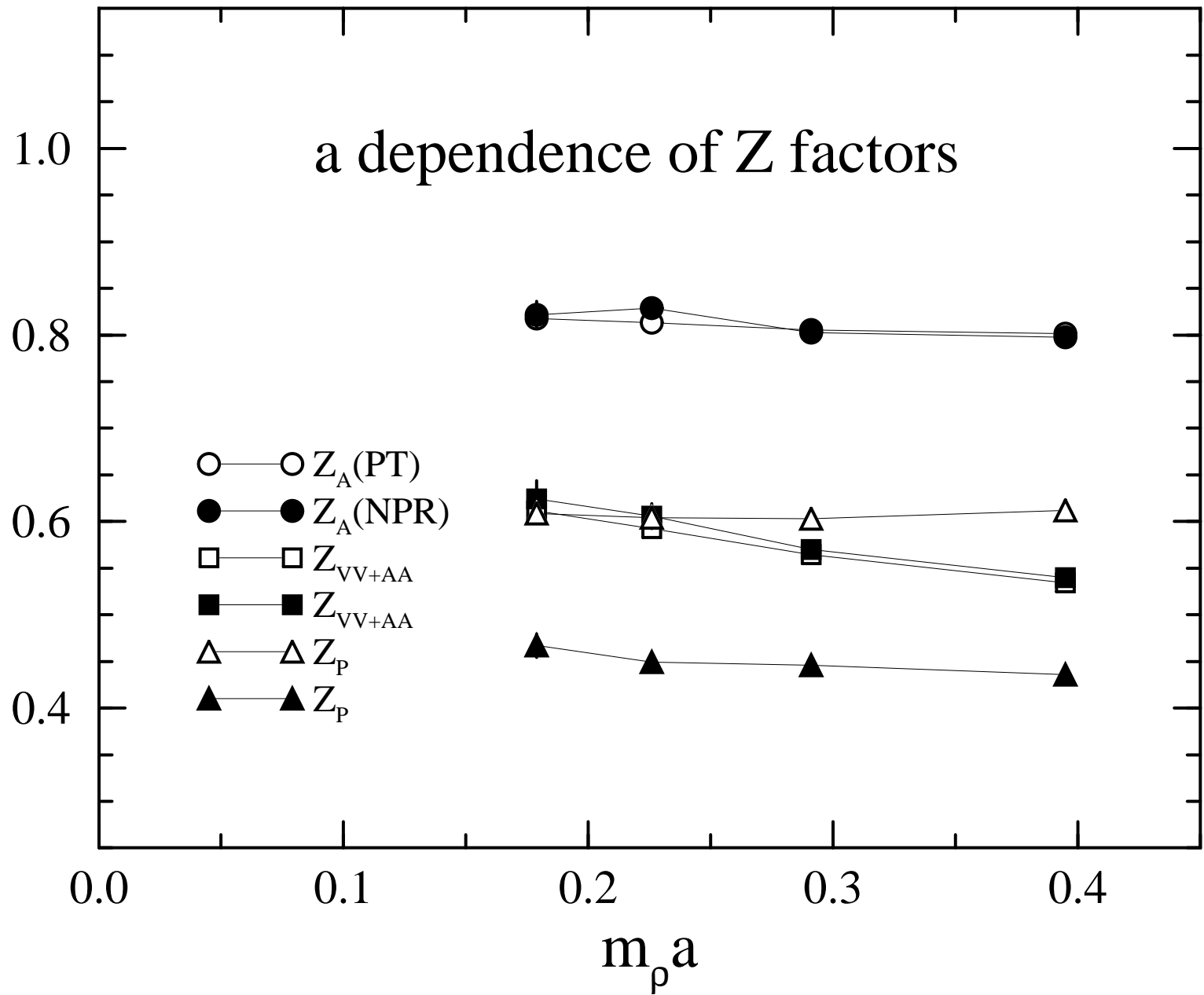
\begin{figure}
\caption{Comparison of renormalization factors
evaluated at ${p^{(*)}}\approx 2$ GeV 
using the NPR (NPR; solid symbols) method  
and the tadpole-improved perturbation theory
(PT; open symbols) 
as a function of $m_\rho a$.}
\label{fig:znpr_beta}
\end{figure}

For completeness let us examine the quark mass
dependence and the $a$ dependence of 
$Z_{VV+AA}$, $Z_A$ and $Z_P$ 
taking their results at $p^{(*)}$ 
(vertical lines in Fig.\ \ref{fig:znpr_63}). 
In Fig.\ \ref{fig:znpr_63_chl}
we plot the NPR results for
$Z_{VV+AA}$, $Z_A$, and $Z_P$ together with tadpole-improved 
perturbative values 
as a function of $m_q=(1/K-1/K_c)/2$ for the case of
$\beta=6.3$.
While we observe little quark mass
dependence for $Z_{VV+AA}$ and $Z_A$,
$Z_P$ clearly decreases as the quark mass decreases.
Figure\ \ref{fig:znpr_beta} shows the $a$ dependence of 
$Z_{VV+AA}$, $Z_A$, and $Z_P$ evaluated at $p^{(*)}$
employing the heaviest quark mass at each $\beta$.
The NPR results for $Z_{VV+AA}$ and $Z_A$ are 
consistent with the perturbative ones at four $\beta$
values, while for $Z_P$ we observe a large deviation 
at each $\beta$. 

\section{Chiral behavior}
\label{sec:result_ch} 

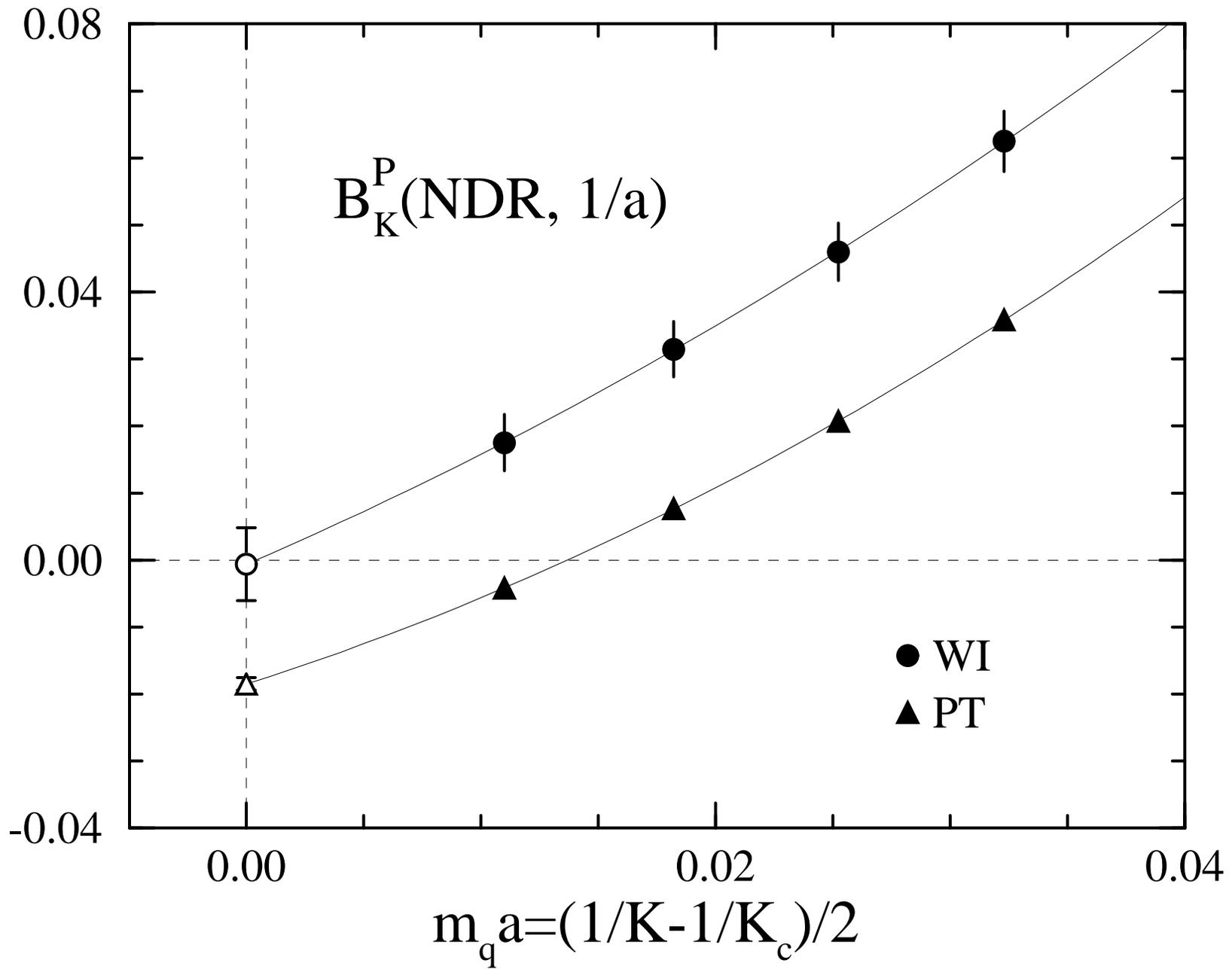
\begin{figure}
\caption{Test of the chiral behavior of  
$B_K^P({\rm NDR},1/a)$ 
for the Ward identity (WI) and perturbative (PT) 
methods at $\beta=6.3$.
Solid curves are quadratic extrapolations to the chiral limit.}
\label{fig:bk_p_63_chl}
\end{figure}

Let us examine the chiral property of the operator 
$\hat \op_{VV+AA}$ using the matrix element $B_K^P$ defined 
in eq.~(\ref{eq:bk_p}), which vanishes in the chiral limit 
in the presence of chiral symmetry.
In Fig.\ \ref{fig:bk_p_63_chl} we show the chiral behavior
of $B_K^P({\rm NDR},1/a)$  
for the case of $\beta=6.3$.
Numerical values of $B_K^P({\rm NDR},1/a)$ for the four 
hopping parameters at each $\beta$ are summarized in
Table~\ref{tab:bk_a}.
The solid lines represent quadratic extrapolations of the WI
and PT results in the bare quark mass $m_q a=(1/K-1/K_c)/2$.
The extrapolated value at $m_q=0$ is consistent with zero, 
demonstrating a significant improvement of the chiral behavior 
compared to the perturbative result plotted with triangles.

Figure\ \ref{fig:bk_p_63_p2} shows the $p^2a^2$ dependence
of $B_K^P({\rm NDR},1/a)$ extrapolated to $m_q=0$, 
which are evaluated at each scale
$p^{(i)}$. Two solid lines
represent the upper and lower bounds of one standard
deviation error for the PT result. 
The values of $B_K^P({\rm NDR},1/a)$ for the WI
method are consistent
with zero within error bars in the momentum range 
$p^2a^2\simlt 1$, albeit the errors become larger 
toward the small momentum region. 

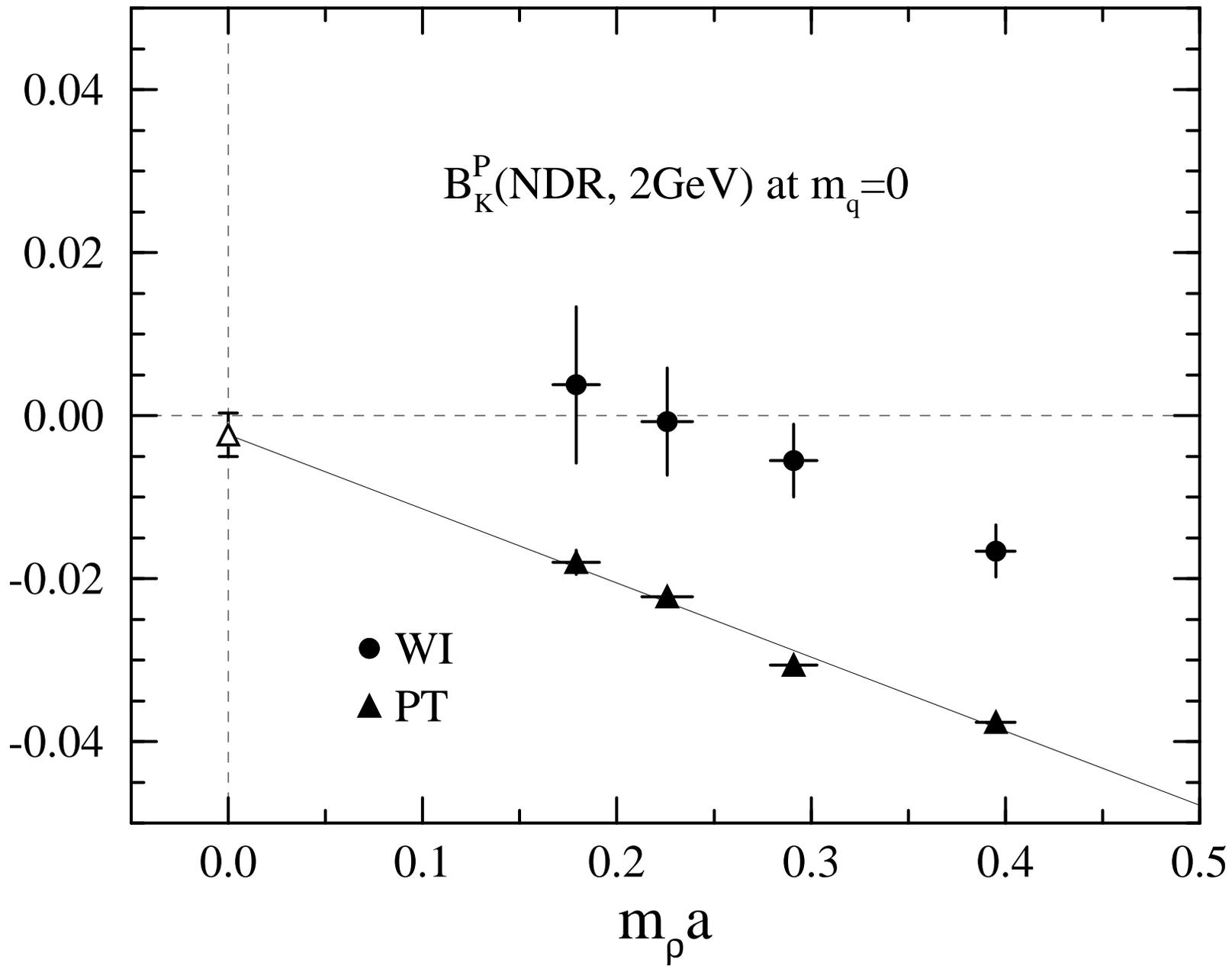
\begin{figure}
\caption{$B_K^P({\rm NDR},2{\rm Gev})$  at $m_q=0$
for the Ward identity (WI) and perturbative (PT) methods 
as a function of $a$.  
Solid line is a linear extrapolation of the perturbative
results to the continuum limit.}
\label{fig:bk_p_beta}
\end{figure}

We plot in Fig.\ \ref{fig:bk_p_beta} the values of 
$B_K^P({\rm NDR},2{\rm GeV})$ in the chiral limit 
as a function of lattice spacing.
The numerical values are given in Table~\ref{tab:bk_2}.
The result for the WI method becomes consistent with zero at
the lattice spacing $m_\rho a\simlt 0.3 (a\simlt 0.08$fm).
In the perturbative approach
with the one-loop mixing coefficients, chiral breaking
effects are expected to appear primarily as terms of $O(a)$ 
and secondly as terms of $O(a\cdot g^2(1/a))$ and
$O(g^4(1/a))$ for the Wilson quark action.  
A roughly linear behavior of our results 
for the perturbative method 
indicate the leading contribution from an $O(a)$ term. 
Making a linear extrapolation 
to the continuum limit $a\to 0$, we observe that 
the chiral behavior is recovered.  
This may suggest that the $O(a\cdot g^2(1/a))$ and
$O(g^4(1/a))$ terms in the mixing 
coefficients left out in the one-loop treatment are 
small or accidentally canceled.


\section{Results for $B_K$}
\label{sec:result_bk} 

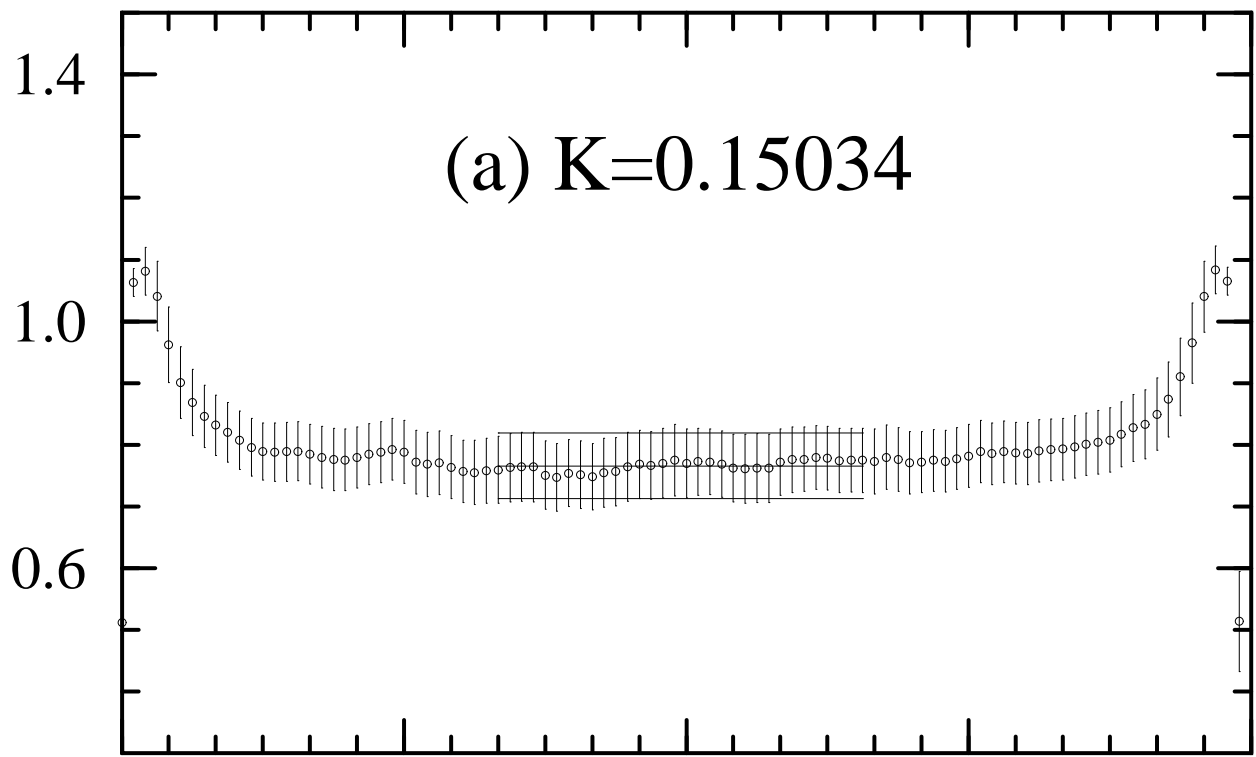
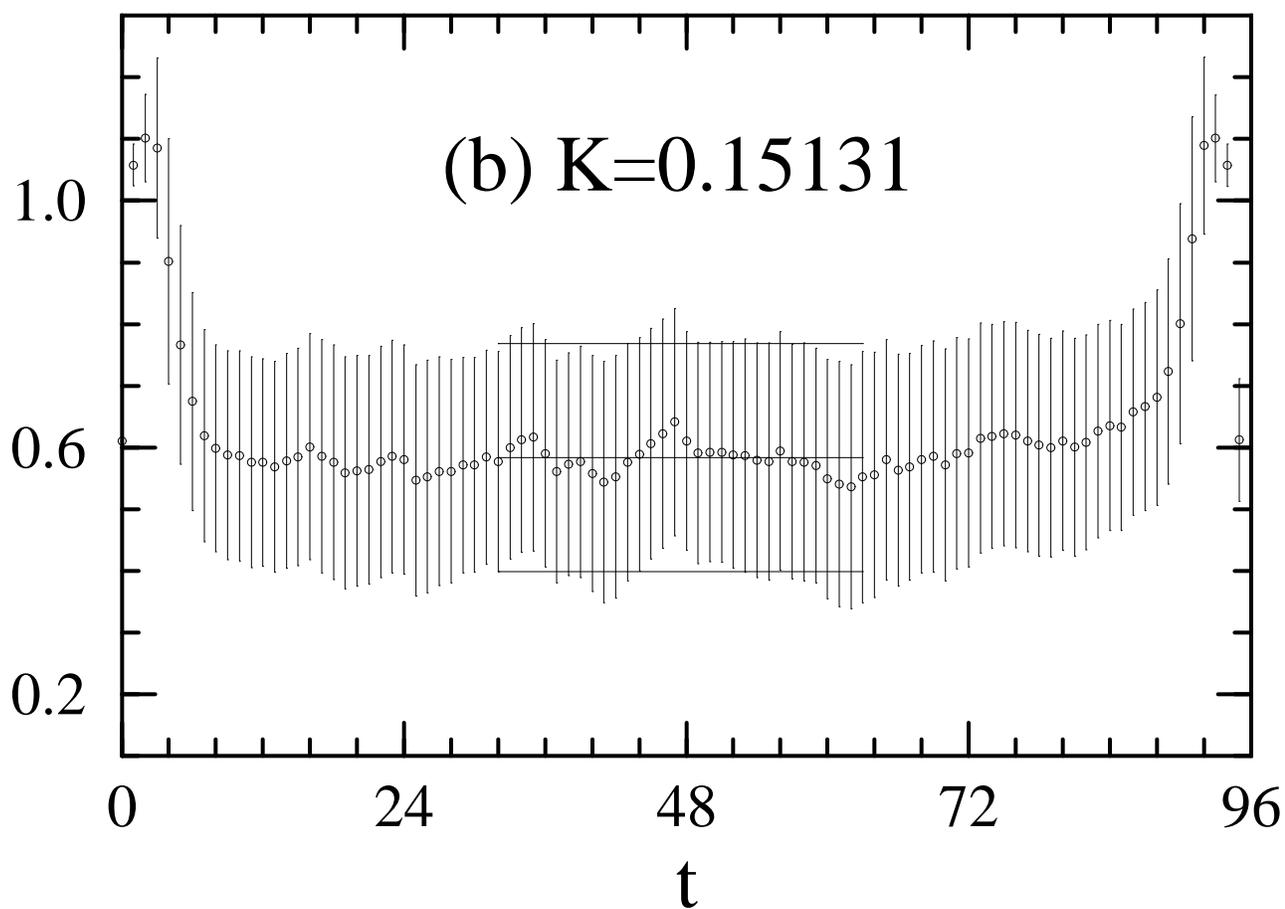
\begin{figure}
\caption{Ratio $R_A(t)$ using mixing coefficients
determined by the Ward identity method
for (a)$K=0.15034$ and (b)$K=0.15131$ 
at $\beta=6.3$. Solid lines denote 
the fitted result and a one standard deviation error band.}
\label{fig:bk_t_63_wi}
\end{figure}

We now turn to the calculation of $B_K({\rm NDR},2{\rm GeV})$.
In Fig.\ \ref{fig:bk_t_63_wi} we present the ratio $R_A(t)$ 
defined in eq.(\ref{eq:ratio}) using the mixing coefficients 
determined from the Ward identities with the external quark
momentum $p^{(*)}$
for the heaviest quark mass($K=0.15034$) and 
the lightest one($K=0.15131$)
at $\beta=6.3$. A good
plateau is observed in the range $20\simlt t \simlt 75$.
We make a global fit of the ratio $R_A(t)$ 
to a constant over $32\le t \le 63$ for this data set.
The three horizontal lines denote the central value 
of $B_K({\rm NDR},1/a)$ and 
a one standard deviation error band.
We note that the error of the fitted
result is roughly  equal in magnitude to those of the ratio over the 
fitted range, while we would usually expect a smaller error for the 
fitted result.
This is because the error of the ratio $R_A(t)$ is
governed by those of the mixing coefficients 
$z_i$ $(i=1,\dots,4)$. Numerical values of 
$B_K({\rm NDR},1/a)$  
for the four hopping parameters at
each $\beta$ are listed in Table~\ref{tab:bk_a}.
  
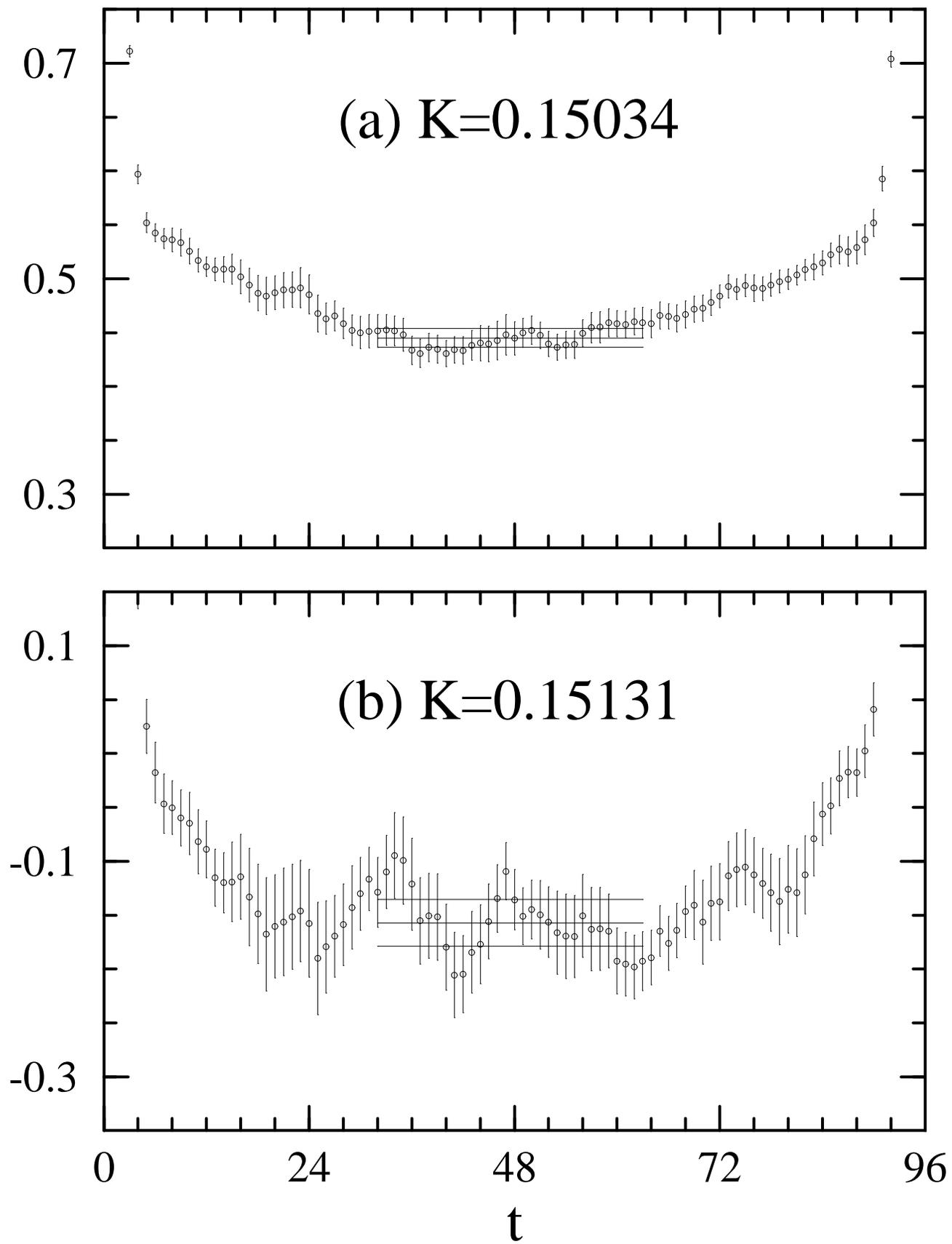
\begin{figure}
\caption{Same as Fig.\ \protect{\ref{fig:bk_t_63_wi}}
for the perturbative method.\hspace{150mm}}
\label{fig:bk_t_63_pt}
\end{figure}

For comparison we also show the results for $R_A(t)$ 
obtained with the perturbative mixing coefficients
in Fig.\ \ref{fig:bk_t_63_pt}. 
We observe a plateau in the range 
$30\simlt t \simlt 65$, which is slightly narrower 
compared to the WI results.
A global fit of $R_A(t)$ to a constant 
choosing the same fitting
range as for the WI case
yields the value of $B_K({\rm NDR},1/a)$
given in Table~\ref{tab:bk_a}.
Let us note that the PT results have quite small errors 
compared to those of the WI method.  This is because 
definite values are taken for the mixing coefficients
in the PT method.

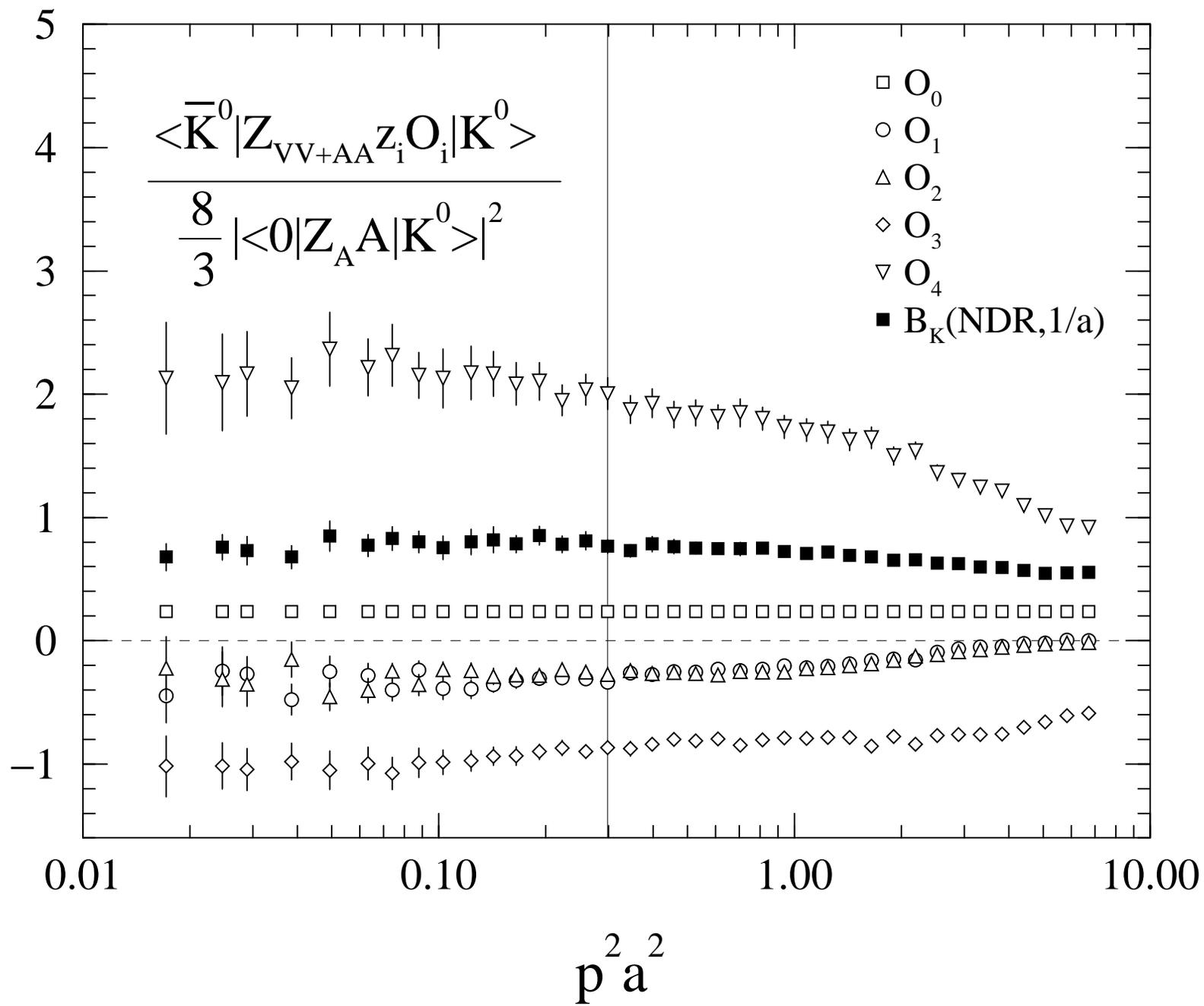
\begin{figure}
\caption{Contributions of the operators 
$\op_i$ $(i=0,\dots,4)$ to $B_K({\rm NDR},1/a)$ with $z_i$
determined by the Ward identity method
for $K=0.15034$ at $\beta=6.3$. Vertical line corresponds to 
$p^{(*)}\approx 2$ GeV.}
\label{fig:bk_63_all}
\end{figure}

In Fig.\ \ref{fig:bk_63_all}
a representative result for 
the contribution of each operator 
$\op_i$ $(i=0,\cdots,4)$ to $B_K({\rm NDR},1/a)$ is shown 
as a function of the external quark momentum
for the case of $K=0.15034$ at $\beta=6.3$, 
which is obtained by fitting the
ratio $R_A^i(t)$ $(i=0,\cdots,4)$ of eq.(\ref{eq:ratio_i}) 
with a constant over the same fitting range 
as for $R_A(t)$. 
The contributions of mixed operators are nearly independent of the
external quark momentum in the 
range $p^2 a^2 \simlt 1.0$. 
An important observation is that 
the value of $B_K({\rm NDR},1/a)$ results from 
large cancellations between the amplitudes 
of the mixing operators $z_i \op_i$ $(i=1,\dots,4)$, 
each having a magnitude 
comparable to or larger than that of $\op_0$.  
This is the essential reason why calculations of $B_K$ with 
the Wilson quark action 
is difficult; the mixing coefficients have to be known accurately 
including higher order effects
both in the coupling constant and the lattice spacing.

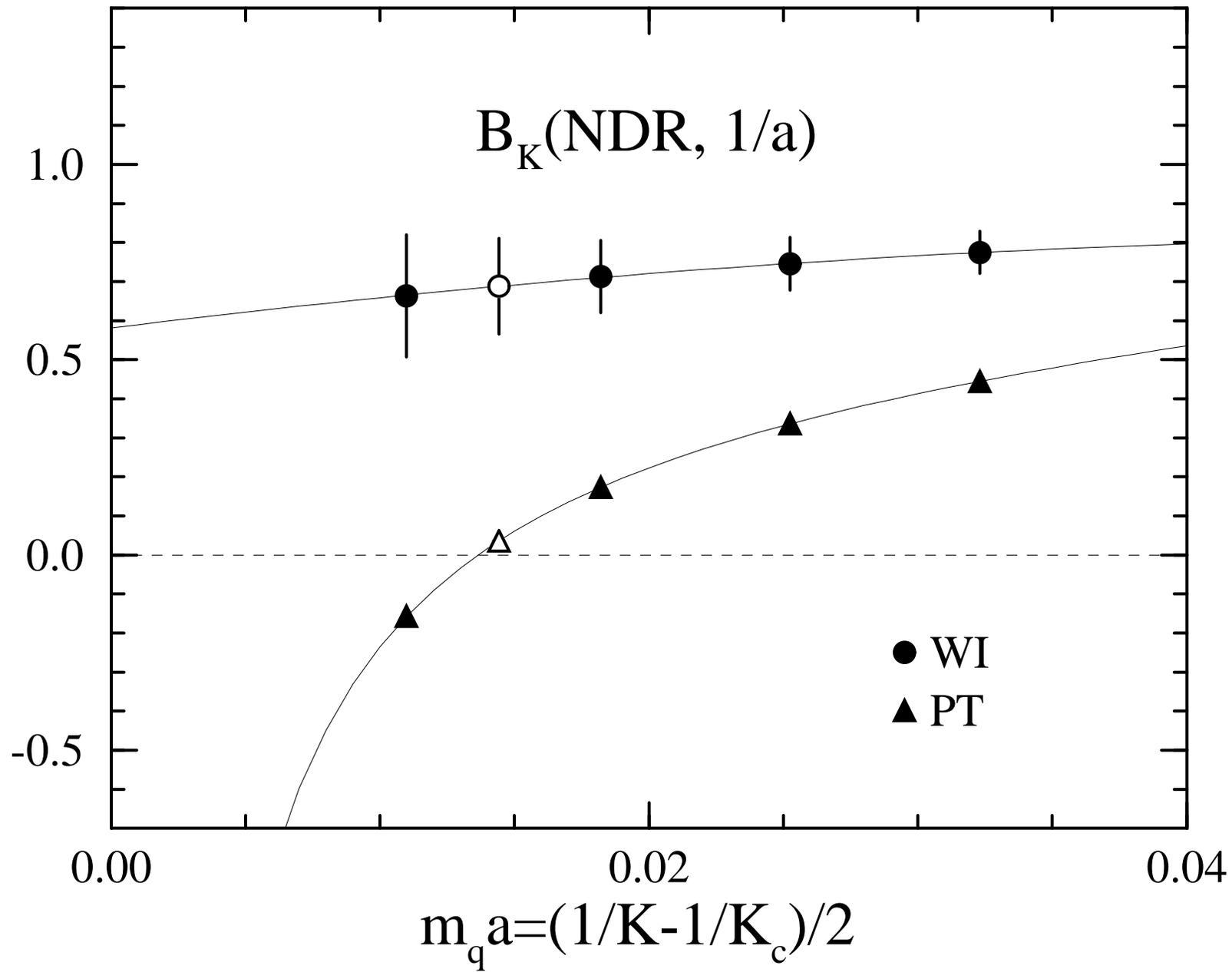
\begin{figure}
\caption{Quark mass dependence of $B_K({\rm NDR},1/a)$
for the Ward identity (WI) method and the tadpole-improved 
perturbative (PT) one at $\beta=6.3$.
Open symbols are interpolations of data 
to $m_s/2$.}
\label{fig:bk_a_63}
\end{figure}

We show the quark mass dependence of $B_K({\rm NDR},1/a)$  
for $\beta=6.3$ in Fig.\ \ref{fig:bk_a_63}.
We observe that the results for the PT method  
seem to diverge toward the chiral limit, while those for
the WI method stay finite.
We expect a different quark mass
dependence for the WI and PT results:
\ben 
B_K({\rm NDR},1/a)&=&A^{\rm WI}+B^{\rm WI} m_q+C^{\rm WI} m_q^2 
\hspace{12pt}{\rm for}\hspace{6pt}{\rm WI}, 
\label{eq:bk_wi_mq} \\
B_K({\rm NDR},1/a)&=&\frac{A^{\rm PT}}{m_q}+B^{\rm PT}+C^{\rm PT} m_q
\hspace{12pt}{\rm for}\hspace{6pt}{\rm PT}, 
\label{eq:bk_pt_mq}
\een
where $A$, $B$, and $C$ are unknown constants. 
These functional forms are based on 
the following assumption for the chiral behavior of the
matrix elements near the chiral limit:
\ben
&&\langle {\bar K}^0 \vert {\hat\op}_{VV+AA} \vert K^0 \rangle
\propto m_q 
\hspace{12pt}{\rm for}\hspace{6pt}{\rm WI}, \\
&&\langle {\bar K}^0 \vert {\hat\op}_{VV+AA} \vert K^0 \rangle
\propto {\rm const} 
\hspace{12pt}{\rm for}\hspace{6pt}{\rm PT}, \\
&&\langle 0 \vert {\hat A}_\mu \vert K^0 \rangle 
\propto \sqrt{m_q}.
\een
For the WI and PT methods we interpolate the data 
at the four hopping parameters 
with the forms (\ref{eq:bk_wi_mq}) 
and (\ref{eq:bk_pt_mq}), respectively, 
to $m_s/2$ and obtain the value of $B_K({\rm NDR},1/a)$ 
at the physical point.

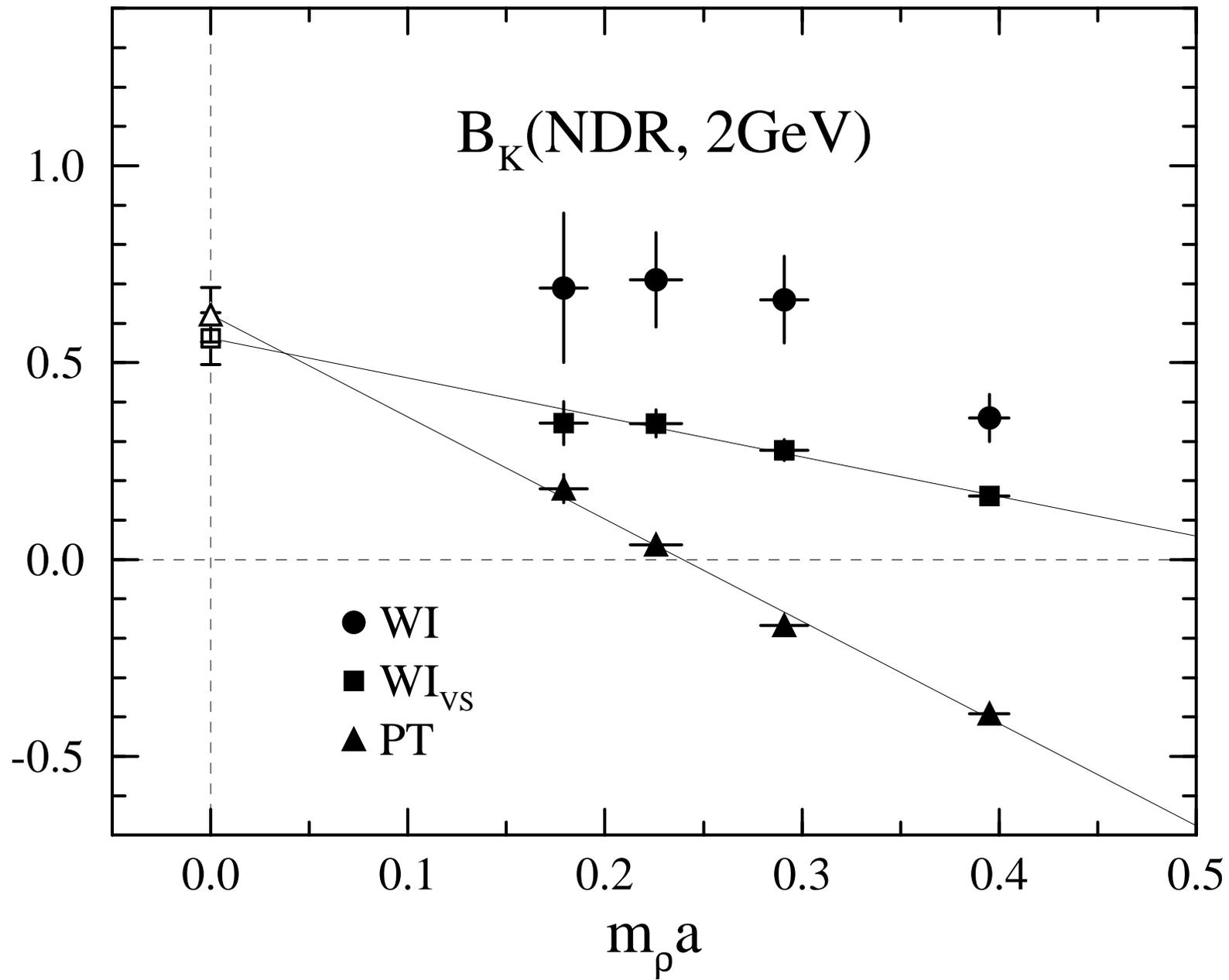
\begin{figure}
\caption{$B_K($NDR, 2GeV) plotted as a function 
of $m_\rho a$ for the WI, WI$_{\rm VS}$ and PT methods. 
Solid lines show linear 
extrapolations to the continuum limit.}
\label{fig:bk_a_cl}
\end{figure}

We summarize our final results for  
$B_K({\rm NDR},2{\rm GeV})$ in Table\ \ref{tab:bk_2}, whose
$a$ dependence is illustrated 
in Fig.\ \ref{fig:bk_a_cl}. 
The method based on the Ward
identity gives a value well convergent from a 
lattice spacing of $m_\rho a\approx 0.3$. 
Unfortunately the large errors do not allow us to take a linear 
extrapolation to the continuum limit.
We may instead take a constant fit of the three results
at smaller lattice spacings($a^{-1}=2.7-4.3$GeV) and find      
$B_K({\rm NDR},2{\rm GeV})=0.68(7)$, which is our best
estimate for the WI method.

Since the origin of the large error is traced to that of the 
mixing coefficients, we attempt to
develop an alternative method, in which the denominator 
of (\ref{eq:Z-factor})
is estimated with the vacuum
saturation of the operator $\hat \op_{VV+AA}$ constructed 
by the WI method:
\be
B_K^{\rm VS}({\rm NDR},1/a)=\frac{\langle
{\bar K}^0 \vert {\hat \op}_{VV+AA} \vert K^0 \rangle}
{Z_A^2 \langle
{\bar K}^0 \vert \op_{VV+AA} \vert K^0 \rangle}_{\rm VS}
\ee
where in terms of eq.(\ref{eq:mix_VV+AA}) the vacuum
saturation of $\hat \op_{VV+AA}$ is rewritten as 
\be
\langle{\bar K}^0 \vert \op_{VV+AA} 
\vert K^0 \rangle_{\rm VS}
=\sum_{i=0}^4  z_i\langle{\bar K}^0 \vert
\op_{i}\vert K^0 \rangle_{\rm VS},
\ee
with
\ben
\langle{\bar K}^0 \vert
\op_{0}\vert K^0 \rangle_{\rm VS}
&=&\frac{8}{3}\langle{\bar K}^0 \vert A\vert 0\rangle
\langle 0\vert A \vert K^0 \rangle, \label{eq:op_0_vs}\\
\langle{\bar K}^0 \vert
\op_{1}\vert K^0 \rangle_{\rm VS}
&=&\frac{8}{3}\langle{\bar K}^0 \vert P\vert 0\rangle
\langle 0\vert P \vert K^0 \rangle, \label{eq:op_1_vs}\\
\langle{\bar K}^0 \vert
\op_{2}\vert K^0 \rangle_{\rm VS}
&=&\frac{4}{3}\langle{\bar K}^0 \vert P\vert 0\rangle
\langle 0\vert P \vert K^0 \rangle, \label{eq:op_2_vs}\\
\langle{\bar K}^0 \vert
\op_{3}\vert K^0 \rangle_{\rm VS}
&=&-\frac{4}{3}\langle{\bar K}^0 \vert A\vert 0\rangle
\langle 0\vert A \vert K^0 \rangle 
-\frac{8}{3}\langle{\bar K}^0 \vert P\vert 0\rangle
\langle 0\vert P \vert K^0 \rangle, \label{eq:op_3_vs}\\
\langle{\bar K}^0 \vert
\op_{4}\vert K^0 \rangle_{\rm VS}
&=&-\frac{8}{3}\langle{\bar K}^0 \vert A\vert 0\rangle
\langle 0\vert A \vert K^0 \rangle 
+\frac{16}{3}\langle{\bar K}^0 \vert P\vert 0\rangle
\langle 0\vert P \vert K^0 \rangle. \label{eq:op_4_vs}
\een
We refer to this as the WI$_{\rm VS}$ method, with which 
the fluctuations in the numerator are expected to largely
cancel against those in the denominator. In fact, errors are
substantially reduced with the WI$_{\rm VS}$ method as
apparent in Fig.\ \ref{fig:bk_a_cl}. 
The cost is that the correct chiral behavior 
of the denominator is not respected  
at a finite lattice spacing due to the contributions of the pseudoscalar
matrix element.
This contribution brings the WI$_{\rm VS}$ result to disagree with
WI at a finite lattice spacing, but the discrepancy should 
vanish in the continuum limit.
A linear extrapolation of the WI$_{\rm VS}$ results in $a$ yields 
$B_K({\rm NDR},2{\rm GeV})=0.562(66)$.

This linear extrapolation, however, involves a systematic uncertainty
arising from the chiral symmetry breaking term 
$c_P|\langle 0 | P|K^{0}\rangle|^2$ in the 
denominator, where $c_P=8/3 z_1+4/3 z_2-8/3 z_3+16/3 z_4$ 
from eqs.(\ref{eq:op_1_vs})$-$(\ref{eq:op_4_vs}).  
Perturbative contributions to $c_P$ starts 
at two-loop order of $O(g^4(1/a))$ as can be checked 
from the one-loop expressions for $z_i$ ($i=1,\dots,4$) 
in eq.(\ref{eq:z_mix_pt}).
Since the matrix 
element $\langle 0 | P|K^{0}\rangle$ diverges in proportion
to $[g^2(1/a)]^{-4/11}$ 
due to the anomalous dimension of the pseudoscalar operator $P$, 
$c_P|\langle 0 | P|K^{0}\rangle|^2$ 
receives contributions of form $[g^2(1/a)]^{14/11}$ which diminishes 
only as a fractional power of $1/\log a$.  
To assess the systematic error associated with this effect, 
we estimate the 
two-loop contribution to $c_P$ by squaring the typical magnitude of 
the one-loop terms in $z_i$: {\it e.g.},   
$|z_i^{one-loop}(\alpha_{\overline{MS}}(1/a))|\simlt 0.08$ at 
$\beta=5.9$ from Fig.\ \ref{fig:zmix_1234_beta}.  
We also estimate 
\be
\frac{\langle 0|P|K^0\rangle }{\langle 0|A|K^0\rangle }
=\frac{m_K}{m_d+m_s} \approx  5
\ee
from the PCAC relation\cite{cppacs},
which yields 
$c_P|\langle 0|P|K^0\rangle |^2/
(8/3)/|\langle 0|A|K^0\rangle |^2\simlt 0.4$.
Since $\alpha_{\overline{MS}}(1/a)^{14/11}$ decreases by 30\% between 
$\beta=5.9-6.5$, over which $a$ decreases by a 
factor 2, this fraction should reduce to $\approx 0.16$ 
after taking the continuum limit.  
Taking account of uncertainties in the choice of coupling constant 
and the mixing coefficients at the two-loop level,
we estimate  
the chiral symmetry breaking contribution of the pseudoscalar density 
that survives after a continuum extrapolation linear in $a$ 
to be $\simlt 20\%$. We conclude
$B_K({\rm NDR},2{\rm GeV})=0.56(7)(11)$ for the WI$_{\rm VS}$ method.

An interesting point in Fig.\ \ref{fig:bk_a_cl} 
is that the perturbative calculation (PT),
which gives a completely ``wrong value'' at $a\ne0$,
yields the correct result for $B_K$, when extrapolated
to the continuum limit $a=0$. 
This is a long extrapolation from negative to positive, 
but the linearly extrapolated value $B_K$(NDR,
2GeV)=0.622(69) 
is reasonable compared with those obtained 
with the WI or WI$_{\rm VS}$ method. This linear
extrapolation is a credible choice because the chiral
behavior of the matrix element
$\langle {\bar K}^0 \vert {\hat \op}_{VV+AA} 
\vert K^0 \rangle$ is linearly recovered as we saw 
in Fig.\ \ref{fig:bk_p_beta}. 
We have to make a reservation, however, 
that this long extrapolation may bring an error larger than 
quoted in the extrapolated value due to systematic effects of
$O(a\cdot g^2(1/a))$ and $O(g^4(1/a))$.
The estimation of these systematic
errors is too complicated
because the matrix elements of the mixing operators
have quite different absolute values.

Each of results from the above three methods suffers from
statistical and systematic errors of $10-20$\% 
which are comparable in magnitude.  
Although the WI$_{\rm VS}$ and the PT methods have the advantage 
of small statistical errors, 
we recognize that this is offset by the difficulty 
to control large systematic errors when attempting a continuum 
extrapolation. 
We thus conservatively take the
result of the WI method $B_K({\rm NDR},2{\rm GeV})=0.68(7)$
at $a^{-1}=2.7-4.3$GeV as our final estimate 
of the present work.

\section{Conclusions}
\label{sec:conclusion} 
 
In this paper we have presented a full account of our method based 
on chiral Ward identities to non-perturbatively determine the 
mixing coefficients of the $\Delta s=2$ operator for the Wilson 
quark action in lattice QCD.  
Implementing the method in a quenched calculation carried out 
at four values of lattice spacing, we have demonstrated 
the effectiveness of the method for constructing
the $\Delta s=2$ operator with the correct chiral property. 
Our final result for $B_K({\rm NDR},2{\rm GeV})=0.68(7)$
at $a^{-1}=2.7-4.3$GeV shows a 
reasonable consistency with 
$B_K({\rm NDR},2{\rm GeV})=0.628(42)$ 
in the continuum limit recently obtained  
with the Kogut-Susskind quark action by us\cite{saoki}.

The error of our Wilson result for $B_K$, however, is still too large to 
convincingly demonstrate that the Wilson and Kogut-Susskind 
quark actions yields the same value in the continuum limit.
We emphasize that this large error is not due to 
an intrinsic defect of the Ward identity method.
It stems from that of large statistical errors 
of the mixing coefficients, which in turn originates 
from our use of point source in evaluating relevant 
quark Green functions. Recent work shows that a variant wall 
source method with the momentum source for the off-shell quark 
propagator\cite{ishizuka,wall_mom} would be effective
to diminish the errors of the mixing coefficients.

Another technical point concerns the issue of Gribov copies 
in the Landau gauge.  While an earlier study\cite{gribov} 
suggests that ambiguities in the choice of the Gribov copies 
induce only small uncertainties comparable to typical 
statistical errors in current numerical simulations, exploring 
gauge invariant implementation of the Ward identity methods, 
either employing the external hadron states or 
the Schr\"{o}dinger functional, 
which is free from this problem, would be worthwhile.

In recent calculations of $B_K$ using 
the $O(a)$-improved quark action
the chiral property of $\hat \op_{VV+AA}$ constructed with
one-loop mixing coefficients shows much improvement
compared to the Wilson quark case\cite{BK_impr}.
This observation can be expected on the ground of
our perturbative results in the Wilson
quark action which suggest 
in Secs.~\ref{sec:result_ch} and \ref{sec:result_bk}  
that the leading
contribution to chiral breaking effects is $O(a)$.
Toward a precise determination of $B_K$ 
the improvement of the quark action is an essential
ingredient.  

A very important physics issue is the effect of quenching. 
With the KS quark action it has been observed that the error
due to quenched approximation is small \cite{bk_ks,OSU}. Whether this
is supported by calculations with Wilson action we must defer to future 
studies. It is straightforward to apply our method once
configurations are generated with dynamical quarks. 

Finally the application of our method
for calculations of $B_B$ would be a worthwhile attempt since previous
calculations of $B_B$ have relied on the mixing
coefficients which were calculated perturbatively in the
massless limit with tadpole improvement.


\acknowledgements

This work is supported by the Supercomputer Project No.32(FY1998)
of High Energy Accelerator Research Organization(KEK), and also 
in part by the Grants-in-Aid of the Ministry of Education 
(Nos. 08640404, 09304029, 10640246, 10640248, 10740107).
Y.K. is supported by Japan Society for Promotion of Science.

\section*{appendix}

The continuum renormalization scheme dependence 
of the renormalization constants $\Delta_{VV+AA}$, 
$\Delta_A$, and $\Delta_P$
defined in eqs.(\ref{eq:z_vv+aa_pt}), 
(\ref{eq:z_a_pt}) and (\ref{eq:z_p_pt}) have been computed 
for the NDR, DRED, and RI schemes by a variety
of authors\cite{z_ndr,z_ri_4} at the one-loop level.  
We consider it useful to reproduce
them in this Appendix using off-shell external
quark states in the general covariant gauge for quark self-energy 
and vertex functions.

We consider the following operators:
\ben
\op_\llop&=&\frac{\delta_{ij}\delta_{kl}
+\delta_{il}\delta_{kj}}{2}
\left({\bar \psi}_1^i
\gamma_\mu(1-\gamma_5)\psi_2^j\right)
\left({\bar \psi}_3^k\gamma_\mu(1-\gamma_5)\psi_4^l\right), \\
\op_{\gamma_\mu\gamma_5}&=&
\delta_{ij}
\left({\bar \psi}_1^i\gamma_\mu\gamma_5\psi_2^j\right), \\
\op_{\gamma_5}&=&\delta_{ij}
\left({\bar \psi}_1^i\gamma_5\psi_2^j\right), 
\een
where $\llop$ represents 
$\gamma_\mu(1-\gamma_5)\otimes\gamma_\mu(1-\gamma_5)$ and
$i$, $j$, $k$, and $l$ label the color indices.
We note that the operator ${\hat \op}_{VV+AA}$ defined 
in eq.(\ref{eq:op_vv+aa})
is parity conserving part of $\op_\llop$ 
with $\psi_1=\psi_3=s$ and $\psi_2=\psi_4=d$.   

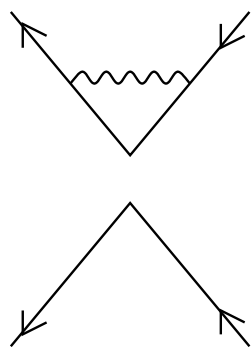
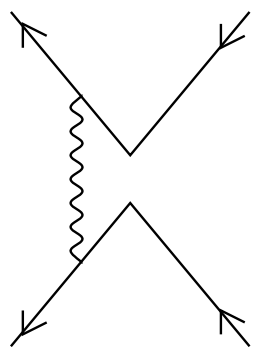
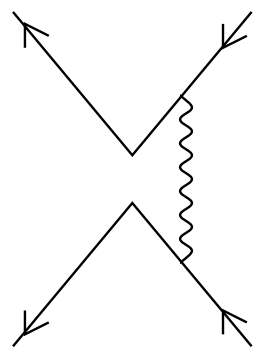
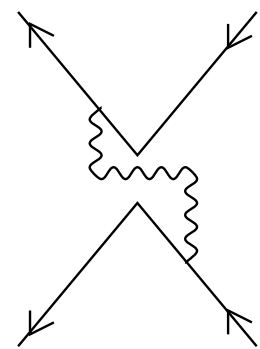
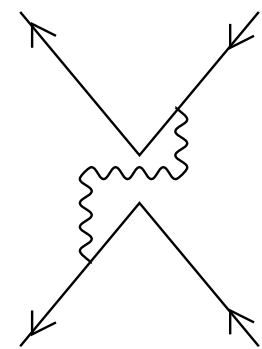
\begin{figure}
\caption{One-loop diagrams for (a) quark self energy,
(b) vertex correction for the quark bilinear operator, and
(c)$-$(h) vertex corrections for the four-quark operator.
$p$ denotes a off-shell momentum for the external quark
state, and $i$, $j$, $k$, and $l$ label color indices.}
\label{fig:ptdgm}
\end{figure}

We draw the relevant one-loop diagrams
in Fig.\ \ref{fig:ptdgm}; (a) the quark self energy, 
(b) the one-loop vertex 
correction for the quark bilinear operators, 
and (c)$-$(h) the six types of the 
one-loop vertex corrections for the four-quark operator. 
The off-shell momentum for the external quark state is denoted as $p$.    
The gauge dependence is parameterized by $\lambda$ 
expressing the gluon propagator as
$\delta_{\mu\nu}/k^2-(1-\lambda)k_\mu k_\nu/k^4$.

Up to the one-loop level the inverse quark propagator and the
vertex functions for $\Gamma=\llop,
\gamma_\mu\gamma_5,\gamma_5$ are written in the following form:
\ben
G^{-1}(p)&=&i\sla{p}{}-\frac{\alpha_s}{4\pi}\Sigma^{(1)}(p),\\
\Lambda_\Gamma(p)&=&\Gamma
+\frac{\alpha_s}{4\pi}\Lambda^{(1)}_\Gamma(p),
\een
where the superscript (i) refers to the $i$-th loop level.
In Table\ \ref{tab:1loop} 
we compile the results for
$\Sigma^{(1)}$ and $\Lambda^{(1)}_\Gamma$  
obtained by employing the NDR scheme 
and the DRED one\cite{dred}. 
The reduced space-time dimension $D$ is
parameterized by $\epsilon$ as $D=4-2\epsilon$, $\epsilon > 0$.
We should note that the one-loop vertex corrections 
yield the extra evanescent
operators which vanish in $D=4$ both for the NDR 
and the DRED schemes. It is meaningless to
give results without mentioning the definition of evanescent
operators, because the constant terms at the one-loop level
depend on the definition of the evanescent operators.
Our choice is as follows:
\ben
E^{\rm NDR}_\llop&=&
\frac{1}{4}\gamma_\rho\gamma_\delta\gamma_\mu(1-\gamma_5)\otimes
\gamma_\mu(1-\gamma_5)\gamma_\delta\gamma_\rho
-\frac{(2-D)^2}{4}\gamma_\mu(1-\gamma_5)\otimes\gamma_\mu(1-\gamma_5),
\label{eq:eop_4_ndr}\\
E^{\rm DRED}_{\gamma_\mu\gamma_5}&=&
{\bar \delta}_{\mu\nu}\gamma_\nu\gamma_5-\frac{D}{4}\gamma_\mu\gamma_5, 
\label{eq:eop_2_dred}\\
E^{\rm DRED}_\llop&=&
{\bar \delta}_{\mu\nu}\gamma_\nu(1-\gamma_5)
\otimes\gamma_\mu(1-\gamma_5)
-\frac{D}{4}\gamma_\mu(1-\gamma_5)\otimes\gamma_\mu(1-\gamma_5),
\label{eq:eop_4_dred}
\een  
where ${\bar \delta}_{\mu\nu}$ is the $D$-dimensional
metric tensor.
 
From the results for
$\Sigma^{(1)}(p)$ and $\Lambda^{(1)}_\Gamma(p)$ $(\Gamma=\llop,
\gamma_\mu\gamma_5,\gamma_5)$ 
we can extract the scheme dependence
of the renormalization constants $\Delta_{VV+AA}$, 
$\Delta_A$, and $\Delta_P$,
which are summarized in Table\ \ref{tab:zdiff}.
For the RI scheme, for which the renormalization constants depend
on the external states and the gauge, we employ the off-shell external 
quark state with momentum $p^2=\mu^2$ and the Landau gauge fixing.

   
\begin{figure}
\caption{Mixing coefficients $z_1,\dots,z_4$ obtained with 
the  Ward identity method
plotted as a function of external momentum squared $p^2a^2$   
for $K=0.15034$ at $\beta=6.3$. Vertical line corresponds to 
$p^{(*)}\approx 2$ GeV.}
\label{fig:zmix_63_ours}
\end{figure}
   
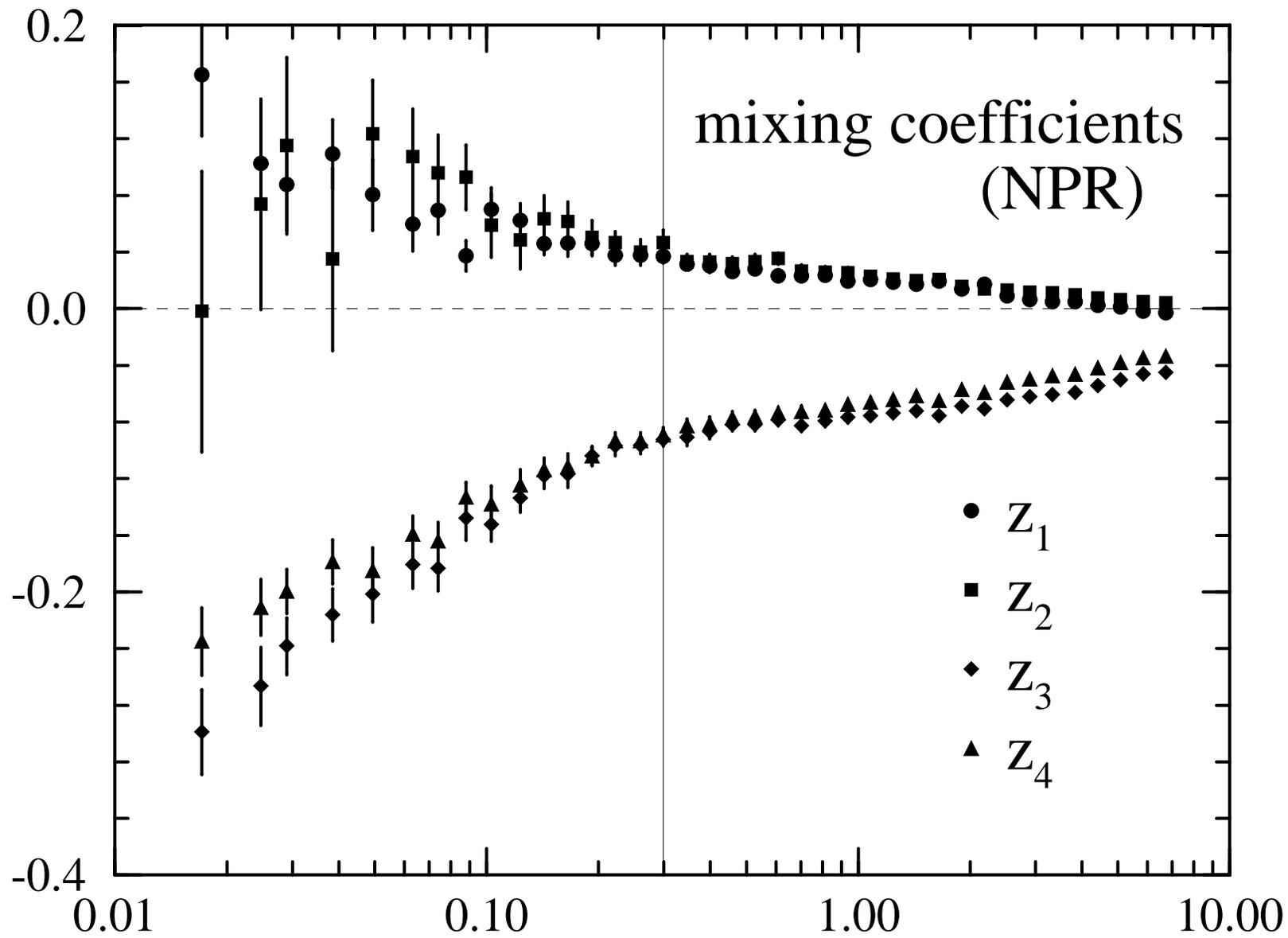
\begin{figure}
\caption{Mixing coefficients $z_1,\dots,z_4$ 
obtained with the NPR method.
Parameters are the same as 
in Fig.\ \protect{\ref{fig:zmix_63_ours}.}}
\label{fig:zmix_63_rome}
\end{figure}
   
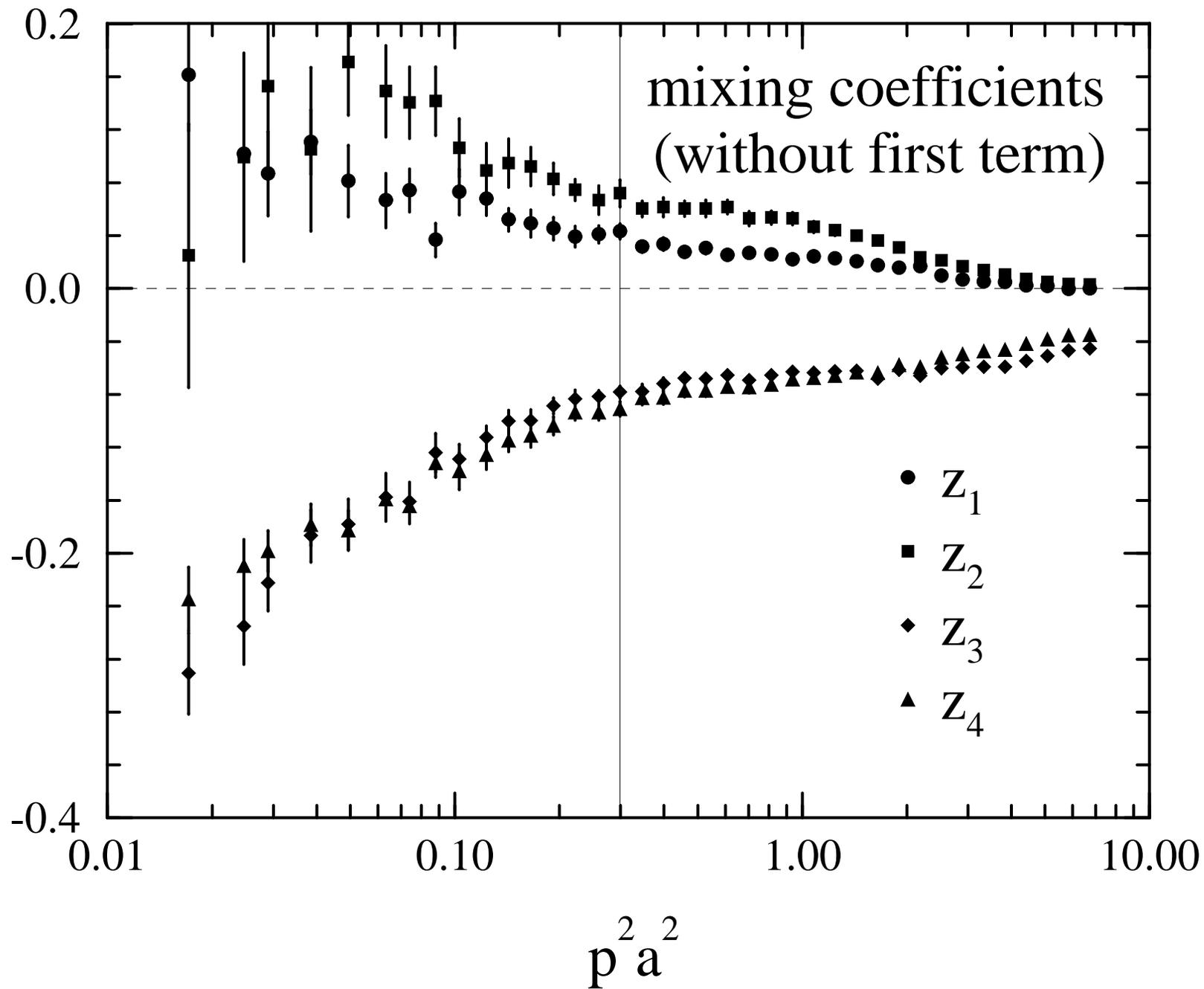
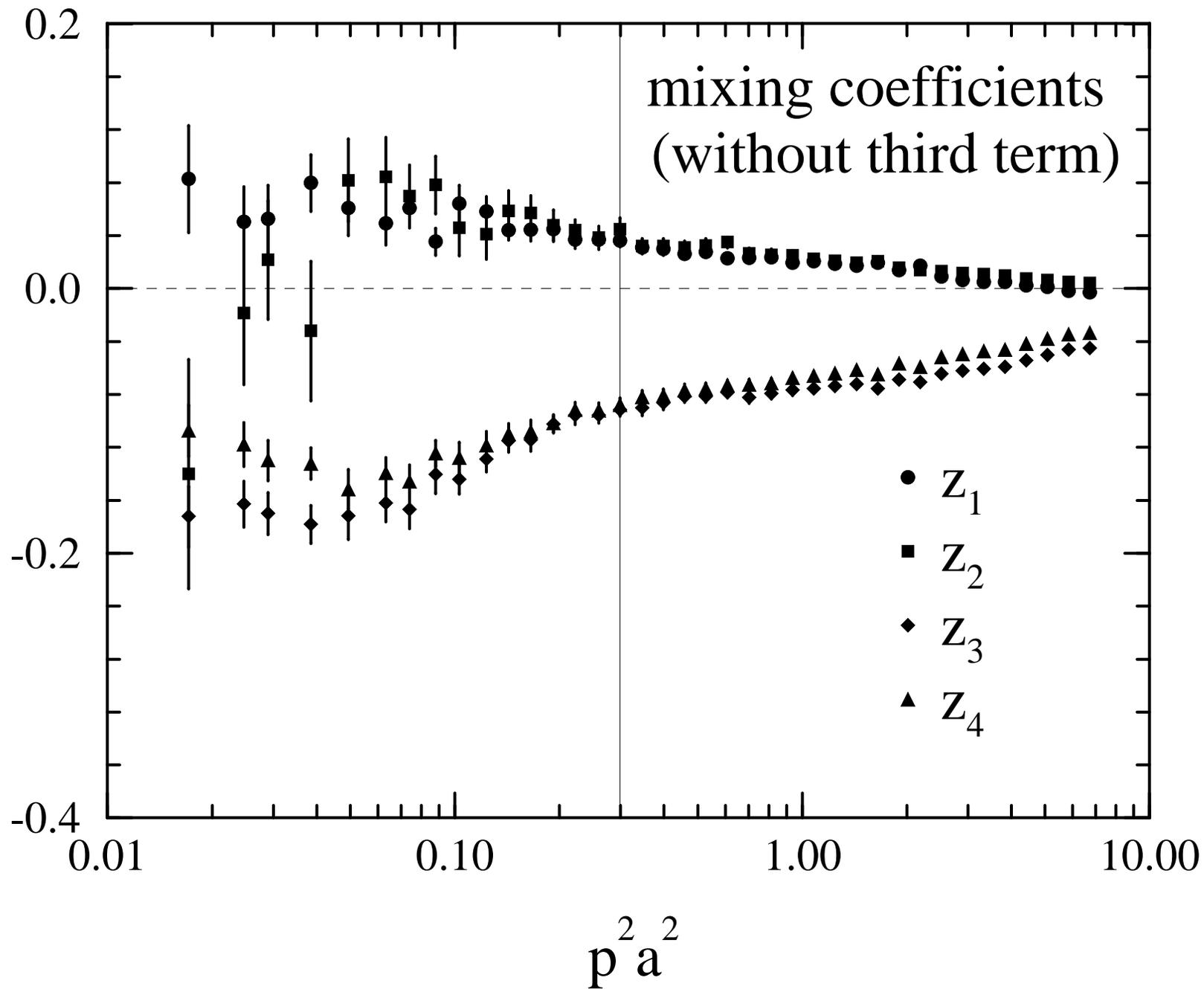
\begin{figure}
\caption{Mixing coefficients $z_1,\dots,z_4$ obtained with 
the Ward identity (\protect{\ref{eq:wi_4qop}}) 
neglecting (a) the first term or (b) the third one. 
Parameters are the same as 
in Fig.\ \protect{\ref{fig:zmix_63_ours}.}}
\label{fig:zmix_63_1234_m50}
\end{figure}

\begin{figure}
\caption{Same as Fig.\ \protect{\ref{fig:zmix_63_ours}}
for mixing coefficient $z_5$.\hspace{150mm}} 
\label{fig:zmix_63_5}
\end{figure}

\begin{figure}
\caption{Dependence of 
$B_K^P({\rm NDR},1/a)$ in the chiral limit 
on the external momentum squared $p^2a^2$ at $\beta=6.3$.
Horizontal solid lines represent the upper and lower bound
of one standard deviation  
error for tadpole-improved perturbative result
in the chiral limit.
Vertical line corresponds to 
$p^{(*)}\approx 2$ GeV.} 
\label{fig:bk_p_63_p2}
\end{figure}

\begin{table}[h]
\begin{center}
\caption{\label{tab:runpara}Parameters of our simulations. See text for 
details.}
\begin{tabular}{lllll}
    $\beta$       & 5.9   & 6.1   & 6.3   & 6.5 \\ 
\hline
$L^3\times T$     & $24^3\times 64$ & $32^3\times 64$ 
                  & $40^3\times 96$ & $48^3\times 96$ \\
No.~conf.         & 300             & 100             
                  & 50              & 24 \\
thermalization    & 22000 & 32000 & 45000 & 72000 \\
interval          & 2000  & 2000  & 5000  & 8000 \\
$K$               & 0.15862  & 0.15428  & 0.15131  & 0.14925 \\ 
                  & 0.15785  & 0.15381  & 0.15098  & 0.14901 \\ 
                  & 0.15708  & 0.15333  & 0.15066  & 0.14877 \\ 
                  & 0.15632  & 0.15287  & 0.15034  & 0.14853 \\ 
$K_c$             & $0.15986(3)$ & $0.15502(2)$     
                  & $0.15182(2)$ & $0.14946(3)$ \\ 
$m_s a/2=m_d a/2$ & $0.0294(14)$ & $0.0198(16)$     
                  & $0.0144(17)$ & $0.0107(16)$ \\ 
$a^{-1}$[GeV]     & 1.95(5)      & 2.65(11)       
                  & 3.41(20)     & 4.30(29) \\ 
$La$[fm]          & 2.4 & 2.4 & 2.3 & 2.2 \\ 
$\langle{\rm Tr}U_P\rangle$
                  & 0.582  & 0.604  & 0.622  & 0.638 \\ 
$\alpha_{\overline{\rm MS}}(1/a)$   
                  & 0.1922  & 0.1739  & 0.1596  & 0.1480 \\ 
$\delta_{p^2}$    & $1.11$ & $1.11$ & $1.15$ & $1.12$ \\
${p^{(*)}}^2a^2$  & $0.9595$ & $0.5012$ 
                  & $0.2988$ & $0.2056$ \\ 
fitting range for $m_\pi,m_\rho,\rho_m,Z_A^{\rm ext}$ 
                  & $12-20$ & $14-24$ & $17-27$ & $20-30$ \\
fitting range for $B_K,B_K^P$
                  & $18-45$ & $24-39$ & $32-63$ & $35-60$ \\
\end{tabular} 
\end{center}
\end{table}

\begin{table}[h]
\begin{center}
\caption{\label{tab:hmass}Meson masses, $\rho_m$ parameter,
and renormalization factor for the extended axial vector
current at 
$\beta=5.9-6.5$ in quenched QCD.}
\begin{tabular}{lllllll}
 $\beta$ & $K$ & $m_\pi$ & $m_\rho$ & $m_\pi$/$m_\rho$ 
& $\rho_m$ & $Z_A^{\rm ext}$ \\ 
\hline
5.9   & 0.15862 & 0.2346(19) & 0.443(10)  & 0.530(12)  
                & 0.03307(33) & 1.328(58)\\ 
      & 0.15785 & 0.2980(15) & 0.4636(49) & 0.6427(67) 
                & 0.05433(32) & 0.941(19) \\ 
      & 0.15708 & 0.3513(12) & 0.4897(33) & 0.7172(45) 
                & 0.07617(31) & 0.925(13) \\ 
      & 0.15632 & 0.3982(11) & 0.5181(25) & 0.7687(33) 
                & 0.09836(29) & 0.919(10) \\ 
6.1   & 0.15428 & 0.1677(16) & 0.323(12)  & 0.520(19)  
                & 0.02239(29) & 0.970(36)\\ 
      & 0.15381 & 0.2135(16) & 0.3467(60) & 0.616(11)  
                & 0.03732(29) & 0.935(23)\\ 
      & 0.15333 & 0.2527(15) & 0.3688(40) & 0.6853(81) 
                & 0.05276(29) & 0.933(17) \\ 
      & 0.15287 & 0.2864(14) & 0.3892(31) & 0.7358(63) 
                & 0.06778(29) & 0.933(14) \\ 
6.3   & 0.15131 & 0.1282(23) & 0.254(14)  & 0.504(27)  
                & 0.01725(34) & 0.981(68)\\ 
      & 0.15098 & 0.1641(20) & 0.2643(65) & 0.621(15)  
                & 0.02889(29) & 0.949(41) \\ 
      & 0.15066 & 0.1933(18) & 0.2797(44) & 0.691(11)  
                & 0.04024(26) & 0.928(32) \\ 
      & 0.15034 & 0.2195(17) & 0.2960(35) & 0.7413(88) 
                & 0.05169(24) & 0.916(28)  \\ 
6.5   & 0.14925 & 0.0782(39) & 0.189(13)  & 0.414(31)  
                & 0.00860(42) & 0.95(14)\\ 
      & 0.14901 & 0.1119(32) & 0.2079(79) & 0.538(21)  
                & 0.01759(42) & 0.951(66) \\ 
      & 0.14877 & 0.1394(29) & 0.2232(60) & 0.625(18)  
                & 0.02658(41) & 0.938(47) \\ 
      & 0.14853 & 0.1632(25) & 0.2368(45) & 0.689(14)  
                & 0.03565(39) & 0.929(37) \\ 
\end{tabular} 
\end{center}
\end{table}

\begin{table}[h]
\begin{center}
\caption{\label{tab:bk_a}$B$ parameters obtained with the
WI, WI$_{\rm VS}$, and PT methods for 
$\beta=5.9-6.5$ in quenched QCD.
Operators are renormalized at $1/a$ in the NDR scheme.}
\begin{tabular}{lllllll}
&& \multicolumn{3}{c}{$B_K({\rm NDR}, 1/a)$} &\multicolumn{2}{c}{
$B_K^P({\rm NDR},1/a)$} \\
 $\beta$ & $K$ & WI & WI$_{\rm VS}$ & PT & WI & PT  \\ 
\hline
5.9   & 0.15862 & 0.270(75) & 0.108(26)  & $-0.631(13)$   
                & 0.0076(21) & $-0.01767(30)$ \\ 
      & 0.15785 & 0.507(39) & 0.259(13)  & $-0.0728(56)$   
                & 0.0236(18) & $-0.00340(25)$ \\ 
      & 0.15708 & 0.620(26) & 0.3617(82) & 0.1937(38)  
                & 0.0408(17) & 0.01274(26) \\ 
      & 0.15632 & 0.687(20) & 0.4348(60) & 0.3501(32)  
                & 0.0585(17) & 0.02977(30) \\ 
6.1   & 0.15428 & 0.62(14)  & 0.226(33)  & $-0.416(24)$   
                & 0.0156(34) & $-0.01039(49)$ \\ 
      & 0.15381 & 0.686(75) & 0.331(20)  & $-0.044(13)$   
                & 0.0293(32) & 0.00186(57) \\ 
      & 0.15333 & 0.720(51) & 0.406(14)  & 0.257(10)  
                & 0.0439(31) & 0.01565(66) \\ 
      & 0.15287 & 0.747(39) & 0.461(11)  & 0.3813(86)  
                & 0.0585(31) & 0.02985(74) \\ 
6.3   & 0.15131 & 0.66(16)  & 0.286(41)  & $-0.157(22)$   
                & 0.0175(42) & $-0.00418(53)$ \\ 
      & 0.15098 & 0.713(92) & 0.387(26)  & 0.173(12)  
                & 0.0315(41) & 0.00765(56) \\ 
      & 0.15066 & 0.745(68) & 0.455(19)  & 0.3362(96)   
                & 0.0460(43) & 0.02072(63) \\  
      & 0.15034 & 0.775(54) & 0.511(16)  & 0.4449(87)  
                & 0.0625(45) & 0.03586(75) \\ 
6.5   & 0.14925 & 0.69(39)  & 0.212(87)  & $-0.338(61)$   
                & 0.0115(65) & $-0.00564(87)$ \\ 
      & 0.14901 & 0.65(17)  & 0.327(46)  & 0.148(25)  
                & 0.0223(59) & 0.00512(92) \\  
      & 0.14877 & 0.67(11)  & 0.406(33)  & 0.329(20)  
                & 0.0348(58) & 0.0171(12) \\ 
      & 0.14853 & 0.699(82) & 0.467(24)  & 0.440(15)  
                & 0.0483(57) & 0.0304(12) \\ 
\end{tabular} 
\end{center}
\end{table}

\begin{table}[h]
\begin{center}
\caption{\label{tab:bk_2}
Results for $B_K($NDR, 2GeV) obtained with the 
WI, WI$_{\rm VS}$ and PT methods 
as a function of $\beta$. Values of $B_K^P({\rm NDR}$,2GeV)  
in the chiral limit for the WI and PT methods are also given.}
\begin{tabular}{lllllll}
    $\beta$       & & 5.9   & 6.1   & 6.3   & 6.5   & $a=0$ \\ 
\hline
$B_K({\rm NDR}$, 2GeV) 
             & WI     
                  & $+0.360(60)$  & $+0.66(11)$ 
                  & $+0.71(12)$   & $+0.69(19)$   & $$ \\
             & WI$_{\rm VS}$     
                  & $+0.162(20)$  & $+0.278(27)$  
                  & $+0.346(35)$  & $+0.347(55)$  & $+0.562(66)$ \\
             & PT
                  & $-0.391(13)$  & $-0.167(20)$   
                  & $+0.037(19)$  & $+0.180(36)$  & $+0.622(69)$ \\
\hline
$\left. B_K^P({\rm NDR},2{\rm GeV}) \right|_{m_q=0}$ 
             & WI
                  & $-0.0166(32)$ & $-0.0055(45)$   
                  & $-0.0007(66)$ & $+0.0038(96)$  & $$ \\
             & PT
                  & $-0.03761(67)$ & $-0.03055(90)$   
                  & $-0.0222(11)$  & $-0.0180(15)$ & $-0.0023(27)$ \\
\end{tabular} 
\end{center}
\end{table}

\begin{table}[h]
\begin{center}
\caption{\label{tab:1loop}
Results for $\Sigma^{(1)}(p)$ and $\Lambda^{(1)}_\Gamma(p)$.
In the entries labeled by ``color'' the color factors are
given with the use of $C_F=4/3$ and $N_c=3$.
In the third column the label F refers to the results in the 
Feynman gauge $\lambda=1$. The label L refers to the
coefficients of the gauge parameter $-(1-\lambda)$ for the 
results in a general gauge. Divergent part is expressed as 
$\pole{1}\equiv 1/\epsilon-\gamma_{\rm E}+\ln{4\pi}$. 
$\sla{p}{}^L\otimes\sla{p}{}^L$ represents 
$p_\mu \gamma_\mu(1-\gamma_5)\otimes 
p_\nu \gamma_\nu(1-\gamma_5)$. Definition of 
evanescent operators $E$ are given 
in eqs.(\protect{\ref{eq:eop_4_ndr}})$-
$(\protect{\ref{eq:eop_4_dred}}).}
\begin{tabular}{llll}
 \multicolumn{4}{c}{(a) NDR}      \\ 
\hline
$\Sigma^{(1)}(p)/(i\sla{p}{})$ 
& \multicolumn{2}{l}{color} & $C_F\delta_{ij}$ \\
& (a) & F  
   & $-\pole{1} -\lnmup -1$ \\     
& (a) & L  
   & $-\pole{1} -\lnmup -1$ \\     
\hline
$\Lambda^{(1)}_{\gamma_\mu\gamma_5}(p)$ 
& \multicolumn{2}{l}{color} & $C_F\delta_{ij}$ \\
& (b) & F  
   & $\gamma_\mu\gamma_5(\pole{1} +\lnmup +1)
     -2 p_\mu\sla{p}{}\gamma_5/{p^2}$ \\     
& (b) & L  
   & $\gamma_\mu\gamma_5(\pole{1} +\lnmup +1)
     -2 p_\mu\sla{p}{}\gamma_5/{p^2}$ \\     
$\Lambda^{(1)}_{\gamma_5}(p)$ 
& \multicolumn{2}{l}{color} & $C_F\delta_{ij}$ \\
& (b) & F  
   & $\gamma_5(\pole{4} +4\lnmup +6)$ \\
& (b) & L  
   & $\gamma_5(\pole{1} +\lnmup +2)$ \\
\hline
$\Lambda^{(1)}_\llop(p)$ 
& \multicolumn{2}{l}{color} 
& $C_F\delta_{ij}\delta_{kl}/2
+(\delta_{ij}\delta_{kl}-\delta_{il}\delta_{kj}/N_c)/4$ \\
& (c), (d) & F  
   & $\llop(\pole{1} +\lnmup +1)-2\llpp$ \\
& (c), (d) & L  
   & $\llop(\pole{1} +\lnmup +1)-2\llpp$ \\
& \multicolumn{2}{l}{color} 
& $(\delta_{il}\delta_{kj}-\delta_{ij}\delta_{kl}/N_c)/4
+(\delta_{ij}\delta_{kl}-\delta_{il}\delta_{kj}/N_c)/4$ \\
& (e), (f) & F  
   & $\llop(-\pole{4} -4\lnmup -9+16\ln{2})
     +E^{\rm NDR}_\llop/\epsilon$ \\
& (e), (f) & L  
   & $\llop(-\pole{1} -\lnmup -2+4\ln{2})$ \\
& \multicolumn{2}{l}{color} 
& $(\delta_{il}\delta_{kj}-\delta_{ij}\delta_{kl}/N_c)/4
+C_F\delta_{il}\delta_{kj}/2$\\
& (g), (h) & F  
   & $\llop(\pole{1} +\lnmup )+2\llpp
     +E^{\rm NDR}_\llop/\epsilon$ \\
& (g), (h) & L  
   & $\llop(\pole{1} +\lnmup )+2\llpp$ \\
\\
 \multicolumn{4}{c}{(b) DRED}      \\ 
\hline
$\Sigma^{(1)}(p)/(i\sla{p}{})$ 
& \multicolumn{2}{l}{color} & $C_F\delta_{ij}$ \\
& (a) & F  
   & $-\pole{1} -\lnmup -2$ \\     
& (a) & L  
   & $-\pole{1} -\lnmup -1$ \\     
\hline
$\Lambda^{(1)}_{\gamma_\mu\gamma_5}(p)$ 
& \multicolumn{2}{l}{color} & $C_F\delta_{ij}$ \\
& (b) & F  
   & $\gamma_\mu\gamma_5(\pole{1} +\lnmup +5/2)
     -2 p_\mu\sla{p}{}\gamma_5/{p^2}
     -E^{\rm DRED}_{\gamma_\mu\gamma_5}/\epsilon$ \\     
& (b) & L  
   & $\gamma_\mu\gamma_5(\pole{1} +\lnmup +1)
     -2 p_\mu\sla{p}{}\gamma_5/{p^2}$ \\     
$\Lambda^{(1)}_{\gamma_5}(p)$ 
& \multicolumn{2}{l}{color} & $C_F\delta_{ij}$ \\
& (b) & F  
   & $\gamma_5(\pole{4} +4\lnmup +8)$ \\
& (b) & L  
   & $\gamma_5(\pole{1} +\lnmup +2)$ \\
\hline
$\Lambda^{(1)}_\llop(p)$ 
& \multicolumn{2}{l}{color} 
& $C_F\delta_{ij}\delta_{kl}/2
+(\delta_{ij}\delta_{kl}-\delta_{il}\delta_{kj}/N_c)/4$ \\
& (c), (d) & F  
   & $\llop(\pole{1} +\lnmup +5/2)-2\llpp 
     -E^{\rm DRED}_\llop/\epsilon$ \\
& (c), (d) & L  
   & $\llop(\pole{1} +\lnmup +1)-2\llpp$ \\
& \multicolumn{2}{l}{color} 
& $(\delta_{il}\delta_{kj}-\delta_{ij}\delta_{kl}/N_c)/4
+(\delta_{ij}\delta_{kl}-\delta_{il}\delta_{kj}/N_c)/4$\\
& (e), (f) & F  
   & $\llop(-\pole{4} -4\lnmup -8+16\ln{2})$ \\
& (e), (f) & L  
   & $\llop(-\pole{1} -\lnmup -2+4\ln{2})$ \\
& \multicolumn{2}{l}{color} 
& $(\delta_{il}\delta_{kj}-\delta_{ij}\delta_{kl}/N_c)/4
+C_F\delta_{il}\delta_{kj}/2$\\
& (g), (h) & F  
   & $\llop(\pole{1} +\lnmup +3/2)+2\llpp
     +E^{\rm DRED}_\llop/\epsilon$ \\
& (g), (h) & L  
   & $\llop(\pole{1} +\lnmup )+2\llpp$ \\
\end{tabular} 
\end{center}
\end{table}

\begin{table}[h]
\begin{center}
\caption{\label{tab:zdiff}
Scheme dependence of renormalization constants 
$\Delta_{VV+AA}$, $\Delta_A$, and $\Delta_P$.}
\begin{tabular}{lll}
           & NDR$-$DRED & NDR$-$RI \\ 
\hline
$\Delta_{VV+AA}$   & $-3$   & $-14/3+8\ln{2}$ \\    
$\Delta_A$         & $-2/3$ & $0$            \\
$\Delta_P$         & $-4/3$ & $16/3$   
\end{tabular} 
\end{center}
\end{table}

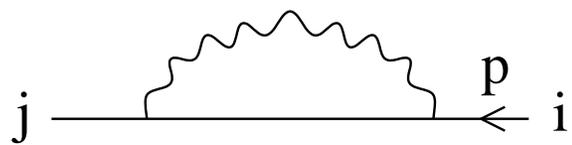
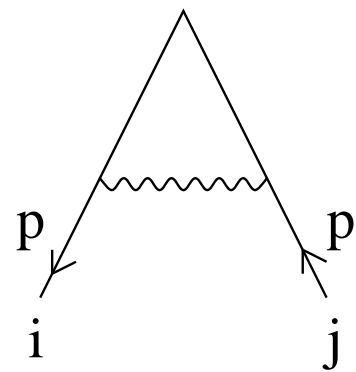
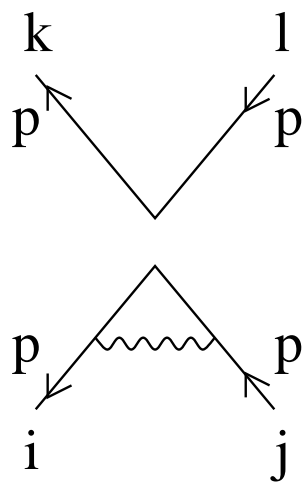
\end{document}